\documentclass[fleqn,twoside]{mystyle}

\usepackage{fancyheadings} 
\usepackage[tbtags]{amsmath}
\usepackage{amssymb}
\usepackage[dvipdfmx]{graphicx}

\usepackage{txfonts}
\def\bm#1{\mbox{\boldmath $#1$}}

\usepackage{cite}

\def\br{\bm{r}}
\def\bp{\bm{p}}
\def\bs{\bm{s}}
\def\bl{\bm{l}}
\newcommand{\rmi}{\mathrm{i}}
\newcommand{\rme}{\mathrm{e}}
\newcommand{\rmd}{\mathrm{d}}
\renewcommand{\Re}{\mathop{\rm Re}}
\renewcommand{\Im}{\mathop{\rm Im}}
\def\<{\langle}\def\>{\rangle}
\def\Tr{\mathop{\rm Tr}}
\def\cE{\mathcal{E}}
\def\rB{\mathop{\rm B}}
\newcommand{\pp}[2]{\frac{\partial #1}{\partial #2}}
\def\etal{\textit{et al} }

\begin{document}

\title{Nuclear shell structures in terms of classical periodic orbits}

\author{%
{\large\bf Ken-ichiro Arita}\\[0.5em]
{\rm Department of Physics, Nagoya Institute of Technology,
Nagoya 466-8555, Japan}\\[0.5em]
{\small E-mail: arita@nitech.ac.jp}
}

\date{Submitted: 31 October 2015,~ revised: 20 April 2016}

\iopabs{
Semiclassical periodic-orbit theory (POT) is applied to the physics of
nuclear structures, with the use of a realistic nuclear mean-field
model given by the radial power-law potential.  Evolution of deformed
shell structures, which are responsible for various nuclear
deformations, are clearly understood from the contribution of short
classical periodic orbits (POs).  Bifurcations of short POs,
which imply underlying local dynamical symmetry, play significant role
there.  The effect of the spin degree of freedom is also investigated
in relevance to the pseudospin symmetry in spherical nuclei and the
prolate-oblate asymmetry in shell structures of nuclei with
quadrupole-type deformations.
}

\maketitle


\section{Introduction}
\label{sec:intro}

Independent-particle picture is one of the most important discoveries
in the history of nuclear structure physics \cite{BM,RS}.  Because of
dominance of the single-particle motion in nuclear dynamics, various
low-energy (near-yrast) properties of nuclei are determined by the
characters of the single-particle spectra.  Therefore, it is very
important to understand the properties of the quantized
independent-particle motion in the nuclear mean-field potentials.  In
general, distribution of the single-particle energy eigenvalues shows
a regularly oscillating gross structure, called \textit{shell
structure} and, occasionally, a modulation in its amplitude into
various kinds of beating patterns called \textit{supershell
structure}.  Those structures are quite sensitive to the shapes of the
potentials.  As is well known, such gross structures in the quantum
fluctuations play important roles in characterizing the properties of
many-fermion systems like nuclei and microclusters.  However, the
origins of such gross structures are not clear from purely quantum
mechanical viewpoints.

Semiclassical theory provides us useful tools to investigate those
remarkable gross shell structures in quantum dynamics.  It describes
the properties of quantum systems in terms of the classical dynamics.
Speaking of a classical-quantum correspondence, one may first think about
the Ehrenfest theorem, which tells that the expectation values of
quantum operators obey the classical equations of motion.  Moreover,
the individual quantum states also have a close relation to the
classical dynamics.  In this relevance, one may recall the
Bohr-Sommerfeld quantization rule for a particle in one-dimensional
potential well $V(x)$, where the energy eigenvalues $\{e_n\}$ are
determined by the condition that the action integral along once around
the classical orbit to be multiples of the Planck's constant $h$:
\begin{equation}
2\int_{x_1}^{x_2}p(x;e_n)\rmd x=nh, \quad n=1,2,\cdots.
\label{eq:BohrSom}
\end{equation}
Here, $p(x;e)=\sqrt{2m(e-V(x))}$ represents the momentum of the
particle moving along the $x$ axis with energy $e$.  The integration
limits $x_1$ and $x_2$ are given by the classical turning points
satisfying $p(x_i;e)=0$.  Semiclassical periodic-orbit theory (POT)
based on the Feynman's path integral formalism has been developed
since 1960s.  Gutzwiller derived the famous \textit{trace formula}
\cite{Gutz} in which the quantum level density (density of states) is
expressed in terms of the classical periodic orbits (POs).  This
formula gives a deep understanding on the origin of quantum
fluctuations.  Even the individual quantum states in non-integrable
systems can be approximately constructed from classical dynamics by
the use of the trace formula, but here attention will be focused on
its important aspects in applications to the gross shell and
supershell structures.

In section~\ref{sec:theory}, semiclassical theories for the
single-particle level densities and for the fluctuations in energies
of many-body systems are briefly outlined.  We will discuss the role
of the classical POs, especially on the importance of the PO
bifurcations and their relation to the restorations of local dynamical
symmetries.  In section~\ref{sec:model}, the radial power-law
potential model and its scaling property are presented.  In
section~\ref{sec:pseudospin}, we apply the semiclassical POT to the
spherical power-law potential model with spin-orbit coupling and study
the origin of the nuclear magic numbers and that of the dynamical
symmetry known as \textit{pseudo-spin symmetry}.  In
section~\ref{sec:exotic}, the nuclear exotic deformations
(superdeformations and octupole deformations) and the roles of PO
bifurcations are analyzed.  In section~\ref{sec:prodom}, the origin of
prolate-shape dominance in nuclear ground-state deformations is
investigated taking the spin-orbit coupling into account.
Section~\ref{sec:summary} is devoted to conclusions and perspectives.

\section{Semiclassical theory of shell structure}
\label{sec:theory}

As one sees from the successes in the mean-field approaches to nuclear
many-body problems, quantum fluctuations in physical quantities are
originated mainly from the shell effect due to the quantized
single-particle motions in the mean field potential.  In this section,
we first describe the Strutinsky shell correction method to extract
\textit{shell energy}, the fluctuation part of the energy for many-body
systems, from the single-particle spectrum.  It will be shown that the
shell energy is expressed in terms of the oscillating part of the
single-particle level density.  Next, we give a brief introduction to
the semiclassical formula for the level density and shell energy,
whose oscillating part is expressed as the sum over contribution of
the classical POs.  We will emphasize the importance
of PO bifurcations for the enhancement of shell effect.

\subsection{Shell correction method}
\label{sec:scm}

In the independent particle picture for an interacting many fermion
system, the constituent particle motion is quantized with the
self-consistent mean-field Hamiltonian.  Particles are arranged to the
quantized states according to the Fermi statistics so that they
minimize the total energy.  Due to the interaction, total energy of
the system considerably differs from a simple sum of the single-particle
energies
\begin{equation}
E_{\rm sp}(N)=\sum_{j=1}^N e_j.
\end{equation}
However, the oscillating part or $E_{\rm sp}$ is found to successfully
describe the energy fluctuation of the total many-particle system.
Strutinsky has derived the way with which one can unambiguously
decompose $E_{\rm sp}$ into the average and oscillating parts
\cite{Strut,BrackPauli} as
\begin{equation}
E_{\rm sp}(N)=\bar{E}_{\rm sp}(N)+\delta E(N).
\label{eq:e_decompose}
\end{equation}
By employing the realistic mean field model and replacing the average
part $\bar{E}_{\rm sp}$ with more reliable
semi-empirical formula, e.g. the liquid drop model (LDM),
\begin{equation}
E(N)=E_{\rm LDM}(N)+\delta E(N),
\end{equation}
one can systematically describe the observed nuclear binding energies
in good precisions.

To calculate the oscillating part $\delta E$ from the
single-particle spectra $e_j$, one first decompose
the single-particle level density
\begin{equation}
g(e)=\sum_j\delta(e-e_j)
\end{equation}
into the average and oscillating parts
\begin{equation}
g(e)=\bar{g}(e)+\delta g(e).
\end{equation}
The average part is obtained by convolving the total one with a
smoothing function $f$, for which a normalized gaussian of width
$\gamma$ with appropriate order of curvature corrections is
usually employed:
\begin{subequations}
\begin{gather}
\bar{g}_\gamma(e)=\frac{1}{\gamma}\int\rmd e' g(e')
 f\left(\frac{e-e'}{\gamma}\right), \\
f(x)=\frac{1}{\sqrt{\pi}}
\rme^{-x^2}L_M^{(1/2)}(x^2).
\end{gather}
\end{subequations}
Here, the $2M$\,th order curvature corrections are given by the Laguerre
polynomial $L_M^{(1/2)}$, and we take $2M=6$ in our numerical calculations.
For a given $\gamma$, the \textit{smoothed} Fermi energies
$\bar{e}_F$ is defined thorough the particle-number condition
\begin{equation}
\int_{-\infty}^{\bar{e}_F(\gamma)}\bar{g}_\gamma(e)\rmd e=N,
\end{equation}
and the average part of (\ref{eq:e_decompose}) is
obtained by
\begin{equation}
\bar{E}_{\rm sp}(N;\gamma)=\int_{-\infty}^{\bar{e}_F(\gamma)}
 e\bar{g}_\gamma(e)\rmd e.
\end{equation}
The smoothing width $\gamma$ is determined so that
$\bar{E}_{\rm sp}$ satisfies the so-called
\textit{plateau condition}
\begin{equation}
\pp{}{\gamma}\bar{E}_{\rm sp}(N;\gamma)\approx 0,
\end{equation}
in order that the obtained $\bar{E}_{\rm sp}(N)$
is less dependent on the physically
meaningless parameter $\gamma$.
Inserting $g(e)=\bar{g}(e)+\delta g(e)$ into the particle-number
condition for the exact Fermi energy $e_F$, one has the relation
\begin{align}
0&=\int_{-\infty}^{e_F}g(e)\rmd e-N \nonumber\\
 &=\int_{-\infty}^{e_F}\{\bar{g}(e)+\delta g(e)\}\rmd e
  -\int_{-\infty}^{\bar{e}_F}\bar{g}(e)\rmd e \nonumber\\
 &=\int_{-\infty}^{e_F}\delta g(e)\rmd e
  +\int_{\bar{e}_F}^{e_F}\bar{g}(e)\rmd e
\label{eq:du_dg}
\end{align}
Shell energy $\delta E$ is then represented in terms of $\delta g$ as
\begin{align}
\delta E &=\int_{-\infty}^{e_F}e\,\left\{\bar{g}(e)+\delta
g(e)\right\}\rmd e
 -\int_{-\infty}^{\bar{e}_F}e\bar{g}(e)\rmd e \nonumber\\
&=\int_{-\infty}^{e_F}e\delta g(e)\rmd e
 +\int_{\bar{e}_F}^{e_F}e\bar{g}(e)\rmd e \nonumber\\
&\approx \int_{-\infty}^{e_F}(e-e_F)\delta g(e)\rmd e.
\label{eq:delta_u}
\end{align}
In the last step, the first $e$ in the integrand of the second term is
replaced with $e_F$, assuming $e_F-\bar{e}_F$ to be sufficiently
small, and then the relation (\ref{eq:du_dg}) is used.  The last
expression (\ref{eq:delta_u}) for the shell energy will be used in the
derivation of its semiclassical formula in the next subsection.

\subsection{Level density in semiclassical approximation}
\label{sec:trace}

In the small $\hbar$ limit, the quantum wave equation reduces to the
classical equation of motion.  Inserting the wave function of the form
$\psi(\br,t)=\rme^{\rmi F(\br,t)/\hbar}$ into the Schr\"{o}dinger equation
$\rmi\hbar\partial\psi/\partial t=\hat{H}\psi(\br,t)$ and expand $F$ in
powers of $\hbar$ as $F=F_0+\hbar F_1+\cdots$, one has the classical
Hamilton-Jacobi equation for $F_0=S$:
\begin{equation}
\pp{S}{t}+H_{\rm cl}(\bp=\bm{\nabla}S,\br)=0
\end{equation}
in the leading order of $\hbar$, where $H_{\rm
cl}(\bp,\br)=\bp^2/2m+V(\br)$ is the classical Hamiltonian.  In the
next-to-leading order, putting $F_1(\br,t)=\frac{1}{2i}\log\rho(\br,t)$,
one has the continuity equation for the probability density $\rho=|\psi|^2$:
\begin{subequations}
\begin{gather}
\pp{\rho}{t}+\bm{\nabla}\cdot(\rho\bm{v})=0, \\
\bm{v}=\frac{\bm{\nabla}S}{m}
\end{gather}
\end{subequations}
and one has a picture of fluid running according to the classical
equations of motion.  In this way, the quantum dynamics can be related
to the classical dynamics in the semiclassical approximation.
Especially, classical POs are shown to play the central
role in the level density and shell energy\cite{Gutz,BaBlo,BBText}.
In the following, we shall briefly outline how the semiclassical
formulas for the level density and shell energy are derived from the
Feynman's path integral representation for the quantum propagator, and
discuss the important aspects of the formulas in analyzing gross shell
structures.

Energy level density $g(e)$ is given by the trace of the
retarded Green's function $G^+(\br'',\br',e)$ as
\begin{gather}
g(e)=\Tr\delta(e-\hat{H})
=-\frac{1}{\pi}\Im\int\rmd\br\, G^+(\br,\br,e), \label{eq:dos} \\
G^+(\br'',\br',e)\equiv\<\br''|\frac{1}{e+\rmi\eta-\hat{H}}|\br'\>,
\end{gather}
where $\eta$ is a positive infinitesimal number.  The Green's function
is given by the Laplace transform of the propagator $K$ as
\begin{gather}
G^+(\br'',\br',e)=\frac{1}{\rmi\hbar}\int_0^\infty \rmd t\,
\rme^{\rmi(e+\rmi\eta)t/\hbar}K(\br'',\br',t), \label{eq:green} \\
K(\br'',\br',t)=\<\br''|\rme^{-\rmi\hat{H}t/\hbar}|\br'\>.
\end{gather}
Connection between quantum and classical mechanics is derived from the
path integral representation for the propagator
\begin{equation}
K(\br'',\br',t)
=\int\mathcal{D}[\br(t)]\exp\left[\frac{\rmi}{\hbar}\int_0^t
 L(\dot{\br},\br)\rmd t'\right], \label{eq:pathintegral}
\end{equation}
where the integral is taken over arbitrary paths connecting initial
point $\br'$ and final point $\br''$ in time $t$.
$\mathcal{D}[\br(t)]$ is the integration measure associated with the
path $\br(t)$, and $L$ is the
Lagrangian function.  The semiclassical
formula of the propagator valid
for small $\hbar$ is obtained by carrying out the above path
integral using
the stationary phase method (SPM).

For an introduction to the basic concept of the SPM,
let us consider a one-dimensional integral of the form
\begin{equation}
I=\int \rmd q\,A(q)\rme^{\rmi S(q)/\hbar},
\end{equation}
with $A(q)$ and $S(q)$ being moderate functions of $q$.  Since $\hbar$
is small, above integrand is a rapidly oscillating function of $q$ and
may have no noticeable contribution to the integral due to the strong
cancellation.  Such cancellation is avoided in vicinity of the
stationary point $q^*$ of the function $S(q)$ satisfying $S'(q^*)=0$,
and it makes a dominant contribution to the integral.  In the standard
SPM, $S(q)$ is expanded around the stationary point $q^*$ up to a
quadratic order, and the above integral is evaluated approximately as
\begin{align}
I &\approx A(q^*)\int_{-\infty}^\infty \rmd q\,
   \exp\left[\frac{\rmi}{\hbar}\left\{
S(q^*)+\tfrac12 S''(q^*)(q-q^*)^2\right\}\right] \nonumber \\
&=A(q^*)\rme^{\rmi S(q^*)/\hbar}\sqrt{\frac{2\pi\rmi\hbar}{S''(q^*)}}.
\label{eq:spa}
\end{align}
In general, $S(q)$ has several stationary points and
equation~(\ref{eq:spa}) will be expressed in the sum over terms
associated with all those points.  This approximation is good for an
isolated stationary point.
However, it becomes worse as the second derivative $S''(q^*)$ becomes
smaller, and then one should consider some higher order expansions of
the action $S(q)$ around $q^*$.

Since the phase in equation~(\ref{eq:pathintegral}) is the action
integral along the path, the stationary solutions are nothing but the
classical trajectory satisfying Hamilton's variational principle.
Then, the propagator is expressed as the sum over contributions of
classical trajectories.  A detailed and clear derivation of the
semiclassical formula from the path integral representation is found
e.g., in section~7 of \cite{BerMou}.  The result is expressed as
\begin{equation}
K_{\rm cl}(\br'',\br',t)=\frac{1}{\sqrt{(2\pi\hbar)^3}}
\sum_{\alpha}\sqrt{D_\alpha}
 \exp\left[\frac{\rmi}{\hbar}R_\alpha-\frac{\rmi\pi\nu_\alpha}{2}
 \right], \label{eq:van-vlech}
\end{equation}
which is known as the Van-Vleck formula.  The sum in the right-hand
side is taken over classical trajectories $\alpha$
starting from $\br'$ and arriving at $\br''$ in time $t$.  $R_\alpha$
represents the action integral along $\alpha$,
\begin{equation}
R_\alpha=\int_0^t L(\br(t'),\dot{\br}(t'))\rmd t',\quad
\br(0)=\br',\quad \br(t)=\br''
\end{equation}
and $D_\alpha$ is given by
\begin{equation}
D_\alpha=\det\left(-\pp{^2R_\alpha(\br'',\br',t)}{\br''\partial\br'}\right)
=\det\left(\pp{\br''}{\bp'}\right)^{-1}
\end{equation}
which is related to the stability of the trajectory with respect to
the initial condition.  $\nu_\alpha$ counts the number of
\textit{conjugate points} along the trajectory $\alpha$, where
the semiclassical propagator encounters singularities in coordinate
space.  Such singularities can be avoided by the Fourier
transformation from the coordinate to momentum space before
it encounters the conjugate point and then inverse Fourier
transformation into coordinate space again after passing
through the point.  This can be also coped with by
the catastrophe theory\cite{SchulBook}.
In such procedure, one generally has the delay of
phase by $\pi/2$, as in the case of one-dimensional WKB wave function
at the classical turning point.

Using the above semiclassical propagator $K_{\rm cl}$ in the Green's
function (\ref{eq:green}) and inserting it into
equation~(\ref{eq:dos}), the level density is expressed in the form
\begin{equation}
g_{\rm cl}(e)=\int \rmd\br \sum_{\alpha}
A_\alpha(\br;e)\exp\left[\frac{\rmi}{\hbar}S_\alpha(\br,\br;e)
-\frac{\rmi\pi}{2}\nu_\alpha\right], \label{eq:g_traceint}
\end{equation}
where the sum is now taken over the closed orbits which start $\br$
with energy $e$ and return to $\br$ again.  $S_\alpha$ is the
Legendr\'{e} transform of the action integral $R_\alpha$, whose
independent variable is transformed from time
$t$ to energy $e=-\partial R_\alpha/\partial t$ as
\begin{align}
S_\alpha(\br'',\br';e)&=et+R_\alpha(\br'',\br',t) \nonumber \\
&=\int_0^t(H+L)\rmd t'=\int_{\br'}^{\br''}\bp\cdot \rmd\br.
\end{align}
For a while, we shall leave out the explicit form of the prefactor
$A_\alpha$ for simplicity, just mentioning that it is related to the
stability of the trajectory with respect to the initial condition.
Finally, the trace integral over $\br$ is carried out with the use of
the SPM.  The stationary phase condition is expressed as
\begin{align}
\pp{S(\br,\br;e)}{\br}
&=\left[\pp{S(\br'',\br';e)}{\br''}+\pp{S(\br'',\br')}{\br'}
\right]_{\br'=\br''=\br} \nonumber \\
&=\bp''-\bp'=0. \label{eq:spcond}
\end{align}
Coincidence of initial and final momenta $\bp''=\bp'$ implies the
orbit to be periodic.  Hence, the semiclassical level density is
expressed in terms of POs as
\begin{equation}
g(e)=\bar{g}(e)+\sum_\beta A_{\beta}(e)\cos\left(
\frac{1}{\hbar}S_\beta(e)-\frac{\pi}{2}\mu_\beta\right).
\end{equation}
The first term $\bar{g}(e)$ represents the average level density which
corresponds to the contribution of zero-length orbit.  The second term
gives the oscillating part of the level density.  The sum is taken
over all the POs (not only primitive ones but also their repetitions).
$S_\beta(e)=\oint_\beta
\bp\cdot \rmd\br$ is the action integral along the orbit $\beta$, and
$\mu_\beta$ is the so-called Maslov phase index related to the
geometric properties of the orbit $\beta$.

The prefactor $A_\beta(e)$ is also expressed in terms of
the classical characteristics of the orbit $\beta$,
such as the stability, period and degeneracy.
Since the action
$S_\beta(e)$ is in general a monotonically increasing function of
energy $e$, each contribution of the PO
gives an oscillating function of $e$, whose
successive minima appear in a distance given by
\begin{equation}
\varDelta e=\frac{2\pi\hbar}{dS_\beta/de}=\frac{2\pi\hbar}{T_\beta}.
\end{equation}
$T_\beta$ represents the period of the orbit $\beta$.  This implies
that the shorter POs having smaller periods $T_\beta$ contribute to
the level density oscillations of larger energy scales (having larger
$\varDelta e$).  Therefore, the gross shell structure is determined by
some shortest POs\cite{StrMag,StrMOD}.  Longer POs
contribute to a finer structure superimposed on the gross one.  Since
the contributions of the POs having different periods give
the terms oscillating with different $\varDelta e$, they will
interfere and build a certain beating pattern.  The supershell
structures, the modulations in shell structures, can be understood as
the result of such interference effect.  Balian and Bloch \cite{BaBlo}
have found a remarkable beating pattern in the coarse-grained level
density for spherical cavity model, and it is understood as the interference
effect of the equilateral triangular and square PO contributions.
Nishioka \etal have employed the semiclassical trace formula to
account for the supershell structures in metallic
clusters\cite{Nishioka}.  They have applied the idea of
Balian and Bloch to a more realistic Woods-Saxon (WS) type
mean field model, and have shown that the supershell structures
observed in metallic clusters are successfully understood as the
interference effect of the triangular and square-type POs.
This is considered as one of the greatest successes in physical
applications of the POT.

Let us next derive the semiclassical expression for
the shell energy $\delta E(N)$.  Inserting the semiclassical level
density
\begin{equation}
\delta g(e)=\sum_\beta A_\beta(e)\cos\left(\tfrac{1}{\hbar}S_\beta(e)
-\tfrac{\pi}{2}\mu_\beta\right)
\end{equation}
into equation~(\ref{eq:delta_u}) and evaluating
the integral using
the semiclassical approximation, one obtains \cite{StrMag,StrMOD}
\begin{align}
&\delta E(N) = \sum_\beta\int_{-\infty}^{e_F}\rmd e(e-e_F)A_\beta(e)
 \cos\left(\tfrac{1}{\hbar}S_\beta(e)-\tfrac{\pi}{2}\mu_{\beta}\right)
 \nonumber\\
&\quad\approx\sum_\beta A_\beta(e_F)
 \int_{-\infty}^{e_F}\rmd e(e-e_F)\,\cos\left[\tfrac{1}{\hbar}
 \left\{S_\beta(e_F) \right.\right.\nonumber\\
&\hspace{11em}\left.\left. +T_\beta(e_F)(e-e_F)\right\}
 -\tfrac{\pi}{2}\mu_\beta\right] \nonumber\\
&\quad =\sum_\beta\left(\tfrac{\hbar}{T_\beta(e_F)}\right)^2A_\beta(e_F)
 \cos\left(\tfrac{1}{\hbar}S_\beta(e_F)
 -\tfrac{\pi}{2}\mu_\beta\right). \label{eq:trace_esh}
\end{align}
From the first to second line in (\ref{eq:trace_esh}), the fact is
used that the integrand in vicinity of the end point $e\approx e_F$
makes the chief contribution to the integral because the integrand is
a rapidly oscillating function of energy due to the smallness of
$\hbar$ and strong offsetting effect arises a little deep inside the
integration region.  Owing to the additional factor
$(\hbar/T_\beta)^2$ in the last expression in
equation~(\ref{eq:trace_esh}), the contribution of longer POs are
relatively suppressed, compared to the trace formula for the level
density, and one usually needs only a small number of the
shortest POs for the study of shell energies.

\subsection{Periodic-orbit bifurcation and local dynamical symmetry}
\label{sec:bif}

\begin{figure}
\begin{center}
\includegraphics[width=.6\linewidth]{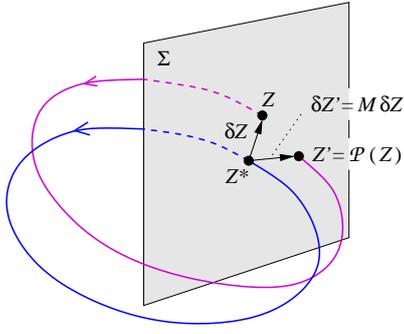}
\end{center}
\caption{\label{fig:poincare}
Illustration of the Poincar\'{e} map $\mathcal{P}(Z)$ defined by the
phase plane $\varSigma$, and the monodromy matrix $M$ associated with
the PO $Z^*$.}
\end{figure}

There are several different ways of deriving PO expansion
formula depending on the integrability of the system.  For a fully
integrable (multiply-periodic) system, it is convenient to use the
action-angle variables $\{\bm{I},\bm{\varphi}\}$, where the action
variables $\bm{I}$ are constants of motion and the Hamiltonian is
independent of the angle variables $\bm{\varphi}$.  In
an $f$-dimensional multiply-periodic system, generic classical
trajectory is confined on a \textit{torus}, an $f$-dimensional
hypersurface in the phase space formed for given values of $\bm{I}$
with varying $\bm{\varphi}$.  In such a system, energy is
quantized according to the EBK (Einstein-Brillouin-Keller) torus
quantization rule\cite{BBText}
\begin{equation}
e_{\{n_k\}}=H(I_k=\hbar(n_k+\tfrac14\alpha_k)),
\quad (n_k=0,1,2,\cdots)\label{eq:ebk}
\end{equation}
which is a generalization of the one-dimensional Bohr-Sommerfeld
quantization rule (\ref{eq:BohrSom}) to multi-dimensional integrable
systems.  Here, $I_k$ is taken as the action integral along the $k$\,th
irreducible loop $\varGamma_k$ on the torus (which cannot be reduced
to a point by any continuous deformation), and $\alpha_k$ is the
so-called Maslov index which counts the caustic points encountered
along $\varGamma_k$.  Based on this quantization rule, Berry and Tabor
derived a formula for the level density expressed as the sum over
terms associated with the classical POs\cite{BerTab}.  Creagh and
Littlejohn have shown that the above Berry-Tabor formula can be also
derived from the phase-space path integral representation of the
quantum propagator\cite{Creagh}.  For partially integrable systems,
some of the trace integrals are carried out exactly, and they bring a
factor proportional to the phase-space volume occupied by the PO
family.  Other integrals are carried out by the SPM, and they
bring a factor related to the stability of the PO \cite{Creagh}.
For a strongly chaotic system in which all the POs are
isolated, one obtains the Gutzwiller trace formula \cite{Gutz,BBText}
\begin{equation}
g(e)=\bar{g}(e)+\sum_\beta
 \frac{T_{\beta}^0(e)}{\pi\hbar\sqrt{|\det(M_\beta-I)|}}
 \cos\left(\frac{1}{\hbar}S_\beta(e)-\frac{\pi}{2}\mu_\beta\right).
\label{eq:gutz_tf}
\end{equation}
Here, $M_\beta$ represents the so-called monodromy matrix which
describes the linearized stability of the PO, and
$T_\beta^0$ represents the period of the primitive PO
$\beta_0$ in case $\beta$ being its repetition.
For a system with $f$ degree of freedoms,
let us consider a $(2f-2)$-dimensional phase plane $\varSigma$
in the $(2f-1)$-dimensional energy surface. This phase plane defines a
stroboscopic mapping known as \textit{Poicar\'{e} map}:  If the energy
surface is compact, the trajectory starting at
the point $Z$ on the phase plane $\varSigma$ will certainly
intersect the same plane again, say, at point $Z'$
with the same orientation as it started off
(see figure~\ref{fig:poincare}).
The map $\mathcal{P}:\varSigma\mapsto\varSigma$ which transforms $Z$
into $Z'$
\begin{equation}
Z'=\mathcal{P}(Z)
\end{equation}
according to the classical trajectory is called Poincar\'{e} map.
PO $Z^*$ is nothing but the fixed point of the
Poincar\'{e} map
\begin{equation}
Z^*=\mathcal{P}(Z^*),
\end{equation}
or more generally,
\begin{equation}
Z^*=\mathcal{P}^n(Z^*)
\end{equation}
which returns to the initial point by the $n$\,th intersection.

The stability of the PO characterizes the behavior of
adjacent trajectories with initial conditions infinitesimally shifted
from $Z^*$.  Expanding the Poincar\'e map around the
PO $Z^*$, the monodromy matrix $M$ is defined by the linear
term as
\begin{subequations}\label{eq:m_monod}
\begin{gather}
Z^*+\delta Z'=\mathcal{P}(Z^*+\delta Z)=Z^*+M\delta Z+O(\delta Z^2),\\
M_{ij}=\pp{Z_i'}{Z_j}. 
\end{gather}
\end{subequations}
The factor $\det(M_\beta-I)$ in equation~(\ref{eq:gutz_tf})
originates from the trace integral in equation~(\ref{eq:dos}) carried
out by the SPM, and this factor is proportional to the curvature of
the action integral
\begin{equation}
C=\det\left(\pp{^2S}{\br\partial\br}\right)_\perp.
\end{equation}
The symbol $\perp$ indicates that the derivatives are taken with
respect to the coordinates perpendicular to the PO, or to the manifold
formed by the PO family under continuous symmetry.  As mentioned
above, the standard SPM breaks down if the curvature $C$ vanishes.
Let us show that a PO \textit{bifurcation} is associated with this
singularity.

\begin{figure}
\begin{center}
\includegraphics[width=\linewidth]{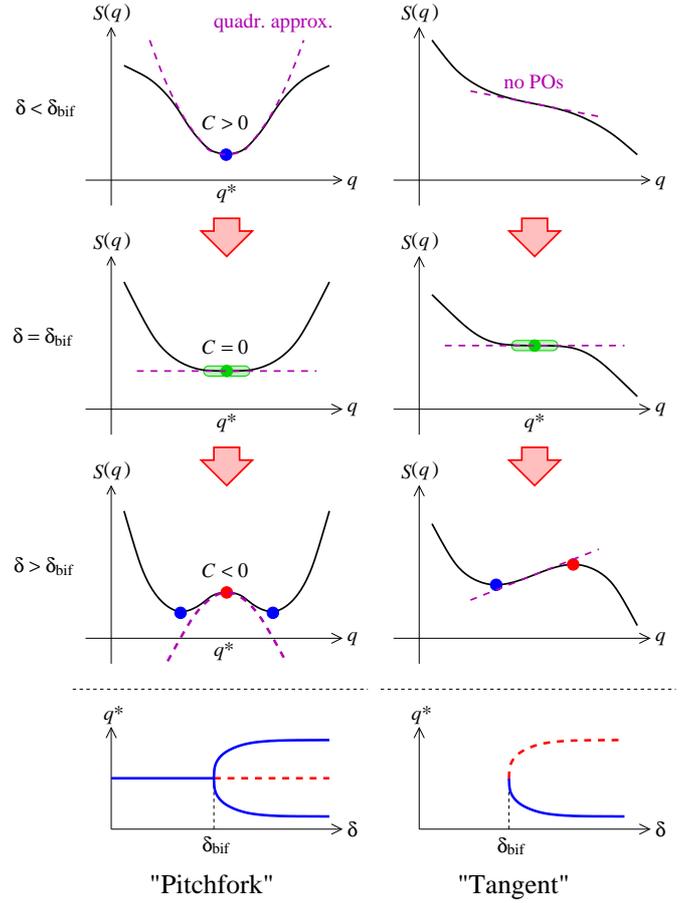}
\end{center}
\caption{\label{fig:bif}
Illustration of PO bifurcation scenarios.  Bifurcations
proceed from the top to lower panels.  POs correspond to the
stationary points $q^*$ of the action $S(q)$, and are indicated by
dots.  The number of POs changes when the parameter $\delta$ passes
over the value $\delta_{\rm bif}$ where the curvature $C=S''(q^*)$
vanishes.  The bifurcations shown in the
left and right columns are called ``pitchfork bifurcation'' and
``tangent bifurcation'', respectively, named after the shapes of the
graphs shown in the bottom panel, where the stationary points are
plotted as functions of the parameter $\delta$.  The inverse processes
(PO association or annihilation) are also
called ``bifurcation'' in a broad sense.}
\end{figure}

Figure~\ref{fig:bif} illustrates two typical scenarios of the
PO bifurcations.  As wee see in
equation~(\ref{eq:spcond}), POs correspond to the
stationary points of the action integral $S(q)$ along the closed orbit
which start from $q$ and returns to $q$ again.  Let us consider the
situation in the top left panel in figure~\ref{fig:bif} where a single
stationary point exists at $q^*$ and the curvature $C$ is positive
there.  One has the factor proportional to $1/\sqrt{C}$ by the
integration with the SPM.  With varying a parameter in the Hamiltonian,
say, the deformation parameter $\delta$, action $S(q)$ will continuously
change and the curvature $C$ may happen to vanish at
$\delta=\delta_{\rm bif}$ as illustrated in the 2nd-row panel.
After passing over this point,
the sign of the curvature $C=S''(q^*)$ changes as
illustrated in the 3rd-row panel, and one has new stationary points at
both sides of the original one.  This is a scenario of the
PO bifurcation which is known as
\textit{pitchfork bifurcation}.  Another type of bifurcation scenario
called \textit{tangent bifurcation} (or \textit{saddle-node
bifurcation}) is shown in the right column of figure~\ref{fig:bif},
where a pair of stable and unstable POs are newly produced at the
``bifurcation'' deformation rather than emerging from already existing
one.\footnote{Change in the number of solutions are generally called
``bifurcation'' in a wide sense.}  All the possible bifurcations in
Hamiltonian systems are classified into six basic types by the
catastrophe theory (see e.g. \cite{Ozorio,OzorioTxt}).

Due to the proportionality $\det(M_\beta-I)\propto C$,
the monodromy matrix $M$ has a unit eigenvalue if the
curvature $C$ vanishes.  This unit eigenvalue suggests the formation
of a local PO family around the bifurcating PO.
If $M$ has a unit eigenvalue, the associated eigenvector
$X_1$ satisfies the relation
\begin{equation}
\mathcal{M}(Z^*+cX_1)\simeq Z^*+cMX_1=Z^*+cX_1,
\end{equation}
where $c$ is a small continuous parameter.  Hence, $Z^*+cX_1$ gives
the continuous family of quasi-periodic family in vicinity of the PO
$Z^*$ as shown in the 2nd-row panels of figure~\ref{fig:bif}.  New
PO(s) may emerge from this family.  The PO bifurcation is thus
associated with a vanishing curvature, or equivalently an emergence of
unit eigenvalue in the monodromy matrix.

The formation of the above local PO family may indicate a
local restoration of dynamical symmetry.
In case where system has a continuous symmetry,
each PO will form a continuous family generated by
the symmetry transformation.  Then, the PO bifurcation
may imply that a dynamical symmetry is locally restored around the
bifurcating PO and generates the
above family of POs around it.
To investigate which kind of invariance is acquired at the
bifurcation point, let us consider the phase-space function
\begin{equation}
D(Z)\equiv\mathcal{P}(Z)-Z.
\end{equation}
It represents the difference of successive intersections on the phase
plane $\varSigma$ by a classical trajectory, and is hence determined
by the Hamiltonian flow.  The PO $Z^*$ is the zero of
above function, namely, $D(Z^*)=0$.  If the monodromy matrix has a unit
eigenvalue and the corresponding eigenvector is $X_1$, one has
\begin{align}
D(Z^* + c X_1)
&=\mathcal{P}(Z^*+c X_1)-(Z^*+c X_1) \nonumber \\
&\simeq cMX_1-cX_1=0
\end{align}
for small continuous parameter $c$.  The local dynamical symmetry is
thus expressed as the invariance of $D(Z)$ around $Z^*$ with respect to the
continuous transformation $Z=Z^*\to Z^*+c X_1$:
\begin{equation}
\left.\pp{D}{X_1}\right|_{Z^*}=0.
\end{equation}

The quasi-periodic family formed around the bifurcating PO is
expected to make a coherent contribution to the path integral, and
brings about a significant shell effect in case it is formed around a
short PO.  Such dynamical symmetry associated with
PO bifurcation sometimes exerts
significant effect on the level statistics\cite{AB2008}.

To examine the effect of the bifurcation on the level
density, it might be useful if
the semiclassical formula valid
also in the vicinity of bifurcation points is available.
The effort of going beyond
the standard SPM to cope with the bifurcation problem
has been made in several approaches.  In the uniform
approximation\cite{Sieber}, action function is expanded up to
appropriate higher order terms.  Those higher order terms have
different function forms depending on the type of bifurcations, and
one has to work out several kinds of catastrophe integrals to obtain
formula valid around those bifurcation points.  In another approach,
the improved SPM\cite{Magner1999,Magner2013} is used
in which the trace integration is carried out by
expanding the phase up to a quadratic order but with keeping the exact
finite integration limits.  These approaches are applied to several
integrable and non-integrable systems and succeeded in reproducing
quantum mechanical results.  In the following sections, we will show
that the shell effects are considerably enhanced
by the effect of the bifurcations
of short POs, which play quite significant roles in
characterizing various nuclear properties.

\section{The radial power-law potential model}
\label{sec:model}

The harmonic oscillator (HO) has been extensively used as a simple
model of the mean-field for qualitative studies of nuclear structures.
It nicely explains the low-energy single-particle spectra for light
nuclei.  It is also useful to understand the appearance of
superdeformed shell structures.  However, heavier nuclei have sharper
potential surface and it is no longer described by the HO model.  To
consider the effect of the sharp surface, modified oscillator model is
devised, in which a term $-v_{ll}\bm{l}^2$ ($\bm{l}$ being orbital
angular momentum vector) is added to the HO potential.  It describes
the effect that the energies of the states with larger $l$, having
major component around the surface, are relatively lowered by
sharpening the potential surface.  Taking also the spin-orbit coupling
term $-v_{ls}\bm{l}\cdot\bm{s}$ into account, this model, known as the
\textit{Nilsson model}, is widely used as a convenient mean
field which provides realistic single-particle levels for
nuclei\cite{Nilsson,BM,RS}.  The square-well potential,
which is further approximated by the infinite-well potential, is also
used for a qualitative description of heavy nuclei.  A realistic
radial profile of the nuclear mean field potential is given by the
WS model having a finite surface diffuseness.  In this
section, we propose a radial power-law potential $V(r)\propto
r^\alpha$ which provides a good approximation to the WS potential and much
easier to treat in both classical and quantum mechanics than the WS
model.  It includes HO and infinite-well models in its two limits
$\alpha\to 2$ and $\alpha\to\infty$, respectively.  We will discuss
the scaling property of the power-law potential model which is
extremely useful in the analysis of both classical and quantum
dynamics.

\subsection{The Hamiltonian and its scaling properties}
\label{sec:scale}

The central part of the Woods-Saxon (WS) potential is
written as
\begin{equation}
V_{\rm c}^{\textsc{ws}}(\br)
=-\frac{V_0}{1+\exp[(r-R_Af(\varOmega;\delta))/a]},
\label{eq:ws_central}
\end{equation}
where $R_A$ and $a$ represent the nuclear mean radius and the surface
diffuseness, respectively.  The shape function $f(\varOmega;\delta)$
describes the angular profile of the nuclear-surface shape with
angle variables $\varOmega=(\theta,\varphi)$ of the spherical coordinate
and the deformation parameter $\delta$.  For sufficiently stable
nuclei, this potential can be approximated by a simpler
power-law (PL) potential
\begin{equation}
V_{\rm c}^{\textsc{pl}}(\br)
=-V_0 + \frac12 V_0\left(\frac{r}{R_Af(\varOmega;\delta)}
\right)^\alpha \label{eq:power_central}
\end{equation}
with a suitable choice of the power parameter $\alpha$.  We determine
the value of $\alpha$ by minimizing the volume integral of the squared
difference of the two spherical potentials $V^{\textsc{pl}}$ and
$V^{\textsc{bp}}$:
\begin{gather}
\frac{\rmd}{\rmd\alpha}
\int_0^{R_A}\rmd r\,r^2\left[V_c^{\textsc{pl}}(r;\alpha)-V^{\textsc{bp}}(r)
\right]^2=0, \\
V^{\textsc{bp}}(r)=-V_0\frac{1+\cosh(R_A/a)}{\cosh(r/a)+\cosh(R_A/a)}.
\end{gather}
Here, we use the Buck-Pilt (BP) potential $V^{\textsc{bp}}$
\cite{Arview,BuckPilt}, which is essentially equivalent to the WS
potential for
surface diffuseness $a$ sufficiently smaller than the radius $R_A$.
The advantage of the BP in contrast with the WS is the absence of singularity
at the origin, which is not critical at the present discussion but
might be
important for the analysis of classical POs intended in the
future.
\begin{figure}
\begin{center}
\includegraphics[width=\linewidth]{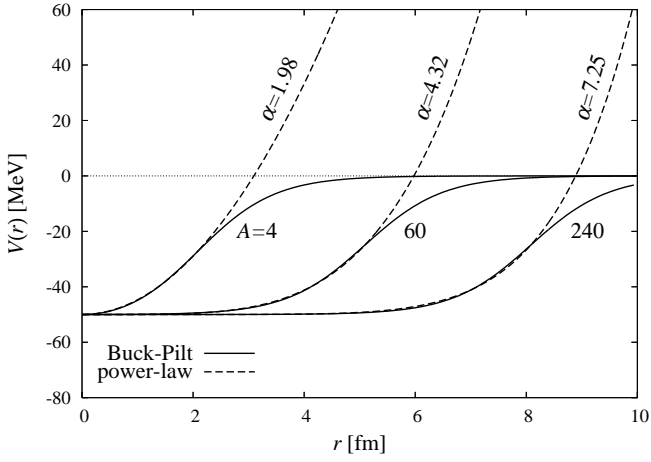}
\end{center}
\caption{\label{fig:ws_alpha}
Radial profile of the radial power-law potential (broken line) fitted to
the Buck-Pilt potential (full line) for several values of mass
number $A$.
Reproduced with permission from \cite{Arita2012}.
Copyright American Physical Society 2012.}
\end{figure}
Figure~\ref{fig:ws_alpha} displays the radial profile of the power-law
potential fitted to the BP potential for several values of the mass
number $A$.  According to the universal WS parameter given in
\cite{Cwiok}, we take the potential depth $V_0=50$~MeV, the radius
$R_A=1.3A^{1/3}$~fm and the surface diffuseness $a=0.7$~fm.
One obtains $\alpha=4\sim 7$ corresponding to the
medium to heavy nuclei $A=50\sim200$.  Removing the constant term in
equation~(\ref{eq:power_central}), we define the model Hamiltonian as
\begin{equation}
H=\frac{p^2}{2m}+U_0\left(\frac{r}{R_0f(\varOmega;\delta)}\right)^\alpha.
\label{eq:h_power}
\end{equation}
Here, $U_0$ and $R_0$ are constants used as the units of energy and
length, respectively, and $m$ is the nucleon mass.  Since the
potential depends on $U_0$ and $R_0$ only in a form $U_0/R_0^\alpha$, $U_0$
and $R_0$ are not necessarily independent and we put $U_0=\hbar^2/mR_0^2$.

Our Hamiltonian (\ref{eq:h_power}) has the following scaling property
\begin{equation}
H(c^{1/2}\bp,c^{1/\alpha}\br)=cH(\bp,\br), \label{eq:scaling}
\end{equation}
regardless of deformation, and the classical equations of motion (EOM)
are invariant under the scaling transformation
\begin{gather}
(\br,\bp,t) \to (c^{1/\alpha}\br,c^{1/2}\bp,c^{1/\alpha-1/2}t)
\end{gather}
with energy $e\to ce$.  This means that if $\br(t)$ is a solution of
EOM at energy $e$, $c^{1/\alpha}\br(c^{1/\alpha-1/2}t)$ gives a
solution of EOM at the energy $ce$.  Therefore, one has the same set of
POs at arbitrary energy, and the action integral along the
orbit $\beta$ is expressed in a simple function of energy as
\begin{equation}
S_\beta(e)=\oint_{\beta(e)} \bp\cdot \rmd\br
=\hbar\tau_\beta\cE.
\end{equation}
The last equation defines dimensionless variables which we call
\textit{scaled period} $\tau_\beta$ and \textit{scaled energy} $\cE$:
\begin{subequations}
\begin{gather}
\tau_\beta\equiv\frac{1}{\hbar}\oint_{\beta(e=U_0)}\bp\cdot \rmd\br, \\
\cE\equiv\left(\frac{e}{U_0}\right)^{1/2+1/\alpha}. \label{eq:scaledvar}
\end{gather}
\end{subequations}
The normal period $T_\beta$ is related to $\tau_\beta$ by
\begin{equation}
T_\beta=\frac{\rmd S_\beta(e)}{\rmd e}
=\hbar\tau_\beta\frac{\rmd\cE}{\rmd e}.
\end{equation}
Then, the Gutzwiller trace formula (\ref{eq:gutz_tf}) for scaled-energy
level density becomes
\begin{align}
&g(\cE)=g(e)\frac{\rmd e}{\rmd\cE} \nonumber \\
&\simeq \bar{g}(\cE)+\sum_\beta
\frac{\tau_\beta}{\pi\sqrt{|\det(I-M_\beta)|}}
\cos\left(\tau_\beta\cE-\tfrac{\pi}{2}\mu_\beta\right).
\label{eq:scaled_gutz}
\end{align}
The average part $\bar{g}$ is given approximately by the
Thomas-Fermi model $g_{\rm TF}$.  For the power-law potential model,
$g_{\rm TF}$ is obtained analytically by
\begin{equation}
g_{\rm TF}(e)=\int\frac{\rmd\bp \rmd\br}{(2\pi\hbar)^3}
\delta(e-H(\br,\bp))
=\frac{1}{\pi\alpha}
\rB\left(\frac{3}{\alpha},\frac32\right)
\frac{\cE^3}{e},
\end{equation}
where $\rB(s,t)$ represents the Euler's beta function.  This average density
is independent of deformation under the volume conservation condition
\begin{equation}
\int \rmd\varOmega f^3(\varOmega;\delta)=4\pi.
\end{equation}
Hence, the average part in equation~(\ref{eq:scaled_gutz}) is given by
\begin{subequations}\label{eq:geps_TF}
\begin{align}
\bar{g}(\cE)\simeq & g_{\rm TF}(e)\frac{\rmd e}{\rmd\cE}=c_0\cE^2,\\
& c_0=\frac{2\sqrt{2}}{\pi}\rB\left(1+\frac{3}{\alpha},\frac32\right).
\end{align}
\end{subequations}

Under the existence of continuous symmetry, POs will be generally
degenerate, namely, they form continuous family generated by the
continuous symmetry transformations.  In a spherical potential,
generic PO forms a three-parameter family generated by the three
independent rotations.  As the exceptions, families of diameter and
circle POs bear only two-parameter degeneracy since they are mapped
onto themselves by one of the rotations.  In an axially-symmetric
potential, generic PO forms a one-parameter family generated by the
rotation about the symmetry axis.  The two exceptions are the diameter
PO along the symmetry axis and the circle PO in the plane
perpendicular to the symmetry axis.  In a system with no continuous
symmetry, all the POs are isolated.  In evaluating the trace integral
with the SPM, one has the additional factor proportional to
$1/\sqrt{S''}\propto\cE^{-1/2}$.  Each continuous symmetry avoids this
factor and hence the contribution of $K$ parameter family has the
energy factor $\cE^{K/2}$ relative to those for
isolated POs.  Taking
account of this energy factor, semiclassical level density of the
power-law potential model is generally expressed as
\begin{equation}
g(\cE)=\bar{g}(\cE)+\sum_\beta A_\beta \cE^{K_\beta/2}
\cos(\tau_\beta\cE-\tfrac{\pi}{2}\mu_\beta),
\label{eq:trace_gen}
\end{equation}
with $A_\beta$ independent of energy.  In systems with continuous
symmetries, there are POs having different degeneracies and $K_\beta$
represents the degeneracy of the family $\beta$.  The derivation of
explicit forms of the amplitude factor under various continuous
symmetries is found, e.g., in \cite{StrMag,Creagh}.

\subsection{Fourier transformation techniques}
\label{sec:fourier}

Due to the simple energy dependence of the action $S_\beta$ in the
power-law potential model, Fourier analysis of the quantum level
density provides us a useful tool to investigate classical-quantum
correspondence.  Let us consider the Fourier transform of the level
density with respect to scaled energy:
\begin{equation}
F(\tau)=\int \rmd\cE g(\cE) \rme^{\rmi\tau\cE} \rme^{-(\gamma\cE)^2/2}.
\label{eq:fourier}
\end{equation}
The last Gaussian factor is introduced for the energy truncation.
With quantum mechanically calculated eigenvalue spectrum
$\{e_j\}$, scaled-energy level density is given by
\begin{equation}
g(\cE)=\sum_j\delta(\cE-\cE_j),\quad
\cE_j=\left(\frac{e_j}{U_0}\right)^{1/2+1/\alpha}. \label{eq:geps}
\end{equation}
Inserting (\ref{eq:geps}) into (\ref{eq:fourier}), one can evaluate
$F(\tau)$ as
\begin{equation}
F^{\rm qm}(\tau)=\sum_j \rme^{\rmi\tau\cE_j}\rme^{-(\gamma\cE_j)^2/2}.
\label{eq:fourier_qm}
\end{equation}
On the other hand, by
inserting the semiclassical level density (\ref{eq:trace_gen})
into (\ref{eq:fourier}),
ignoring the energy dependence of the amplitude for simplicity,
one obtains the semiclassical expression
\begin{equation}
F^{\rm cl}(\tau)=\bar{F}(\tau)
+\pi \sum_\beta A_\beta \rme^{-\rmi\pi\mu_\beta/2}
 \delta_\gamma(\tau-\tau_\beta).
\label{eq:fourier_cl}
\end{equation}
Here, $\delta_\gamma(x)$ represents the normalized Gaussian with width
$\gamma$.  Equation~(\ref{eq:fourier_cl}) tells that $F(\tau)$ is a
function having successive peaks at the scaled periods of classical
POs $\tau=\tau_\beta$ with the corresponding heights
proportional to the amplitude $A_\beta$.
Thus, one can
extract information on the contribution of classical POs
to the level density out of the quantum Fourier transform
(\ref{eq:fourier_qm}).  The present method is very useful in
examining classical-quantum correspondence, especially when the
semiclassical amplitudes are difficult to obtain due to the hidden
(exact or approximate) symmetries, bifurcations and so on.  To obtain
finer resolution (small $\gamma$) of POs in the Fourier spectrum,
quantum spectra up to higher energy ($\cE\sim 1/\gamma$) is required
in evaluating (\ref{eq:fourier_qm}).

\subsection{Spin-orbit coupling}
\label{sec:ls-coupling}

It is well known that the nuclear mean field potential has a strong
spin-orbit coupling.  In the WS model, the spin-orbit term
\begin{gather}
\frac{\lambda}{2(mc)^2}
\left[\bm{\nabla}\frac{V_0}{1+\exp\{(r-R_Af(\varOmega;\delta))/a\}}
\right]\cdot(\bs\times\bp)
\label{eq:ws_so}
\end{gather}
is added to the central potential.
In the same manner as above, we introduce the spin-orbit term in our
power-law potential model as
\begin{equation}
H=\frac{\bp^2}{2m}+U_0\left(\frac{r}{R_0f(\varOmega;\delta)}\right)^\alpha
+2\kappa\left[\bm{\nabla}V_{\rm so}(\br)\right]\cdot(\bs\times\bp),
\label{eq:h_ls}
\end{equation}
with the spin-orbit potential
\begin{equation}
V_{\rm so}(\br)=\frac{1}{m}\left(\frac{r}{R_0f(\varOmega;\delta)}
 \right)^{\alpha_{\rm so}}.
\end{equation}
Although the spin-orbit potential almost equivalent to the central one
is used in the WS model, it might not be so bad to use the power parameter
$\alpha_{\rm so}$ a little different from $\alpha$
in the central potential (\ref{eq:power_central}).  Here we choose
$\alpha_{\rm so}=1+\alpha/2$ in order to keep the scaling relation
\begin{equation}
H(c^{1/2}\bp,c^{1/\alpha}\br,\bs)=cH(\bp,\br,\bs).
\end{equation}
Apparently, this scaling is effective only for \textit{frozen-spin}
motions
where spin vector is static.  Fortunately, spin is frozen in
many important POs, and the above scaling
turns out to be very useful in our semiclassical analysis.  The
spin-orbit coupling strength $\kappa$ is determined so that the
spin-orbit potential in the spherical limit takes the same value as
that of the realistic WS model at the nuclear surface $r=R_A$.
The potential parameters obtained for several values of
mass number $A$ are
shown in table~\ref{tab:params}.
We obtain $\kappa=0.05\sim 0.06$ $\alpha=5.0\sim 6.0$ for medium-mass
region $A=50\sim150$.

\begin{table}
\caption{\label{tab:params}
The values of the parameters $\alpha$, $R_0$, $U_0=\hbar^2/mR_0^2$ and
$\kappa$ in (\ref{eq:h_ls}), obtained by fitting to the WS/BP model
for several values of the mass number $A$.}
\begin{center}
\begin{tabular}{c|c|c|c|c} \hline\hline
$A$ & $\alpha$ & $R_0$~[fm] & $U_0$~[MeV] & $\kappa$ \\ \hline
~20 & 2.80 & 2.32 & 3.32 & 0.089 \\
100 & 5.23 & 3.93 & 1.14 & 0.059 \\
200 & 6.75 & 5.06 & 0.72 & 0.049 \\ \hline\hline
\end{tabular}
\end{center}
\end{table}

The EOM for the classical spin variables are
derived by the spin coherent-state path integral method
\cite{KuraSuz}.  One useful choice of the canonical variables for the spin
degree of freedom is $q_s=\varphi$ and $p_s=s\cos\vartheta$, where
$\vartheta$ and $\varphi$ are polar and azimuthal angles in the spherical
spin coordinates, respectively.  The Cartesian spin components are
given by
\begin{subequations}
\begin{align}
s_x&=s\sin\vartheta\cos\varphi, \\
s_y&=s\sin\vartheta\sin\varphi, \\
s_z&=s\cos\vartheta,
\end{align}
\end{subequations}
with the modulus $s$ constant ($s=\hbar/2$ for nucleon).  One can prove
the Poisson bracket relation between the classical spin variables
\begin{equation}
\{s_i,s_j\}_{\rm P.B.}=\pp{s_i}{q_s}\pp{s_j}{p_s}
-\pp{s_i}{p_s}\pp{s_j}{q_s}=\epsilon_{ijk}s_k,
\end{equation}
which exactly corresponds to the commutation relation of the quantum
spin operators.  The trace formula in extended phase space
including the spin degree of freedom is formulated in \cite{PlecZait}.

Writing $\bm{B}=\bm{\nabla}V_{\rm so}(\br)\times\bp$, classical EOM is
expressed as
\begin{subequations}
\label{eq:eom_spin}
\begin{gather}
\dot{\br}=\pp{H}{\bp}=\frac{\bp}{m}-2\kappa
(\bm{s}\times\bm{\nabla}V_{\rm so}),\label{eq:eom_spin_r} \\
\dot{\bp}=-\pp{H}{\br}=-\bm{\nabla}V_c-2\kappa\bm{\nabla}(\bm{B}\cdot\bs),
\label{eq:eom_spin_p} \\
\dot{\bs}=\{\bs,H\}_{\rm P.B.}=-2\kappa\bm{B}\times\bs.
\label{eq:eom_spin_s}
\end{gather}
\end{subequations}
Let us consider the case where the potentials $V$ ($V_{\rm c}$ and
$V_{\rm so}$) are axially symmetric.
Frozen-spin orbits appear under the
following conditions:
\begin{enumerate}
\item \textit{Meridian and equatorial orbits} \\
Taking $z$ axis as the symmetry axis of rotation, consider a classical
trajectory starting with $\br$ and $\bp$ both on the meridian plane
(the plane containing the symmetry axis), say, the $(x,z)$ plane, and
$\bs$ perpendicular to it, namely, in the $y$ direction.  On the
$(x,z)$ plane, $\bm{\nabla}V$ is perpendicular to the $y$ axis and
then the vector $\bm{B}$ is parallel to the $y$ axis.  Consequently,
the $y$-components of all the terms in the right-hand sides of
equations~(\ref{eq:eom_spin_r}) and (\ref{eq:eom_spin_p}) as well as
the right-hand side of (\ref{eq:eom_spin_s}) vanish, and the
trajectory is shown to remain in the $(x,z)$ plane with its spin
frozen.  Hence one has the meridian-plane frozen-spin orbits.  If the
potential is also symmetric with respect to the $(x,y)$ plane
(equatorial plane), the classical orbits in this plane with spin
perpendicular to it are shown just as above to be frozen-spin orbits.
\item \textit{Diameter orbits} \\
Consider a trajectory starting along the symmetry axis ($z$ axis) with
spin parallel to the $z$ axis.  On the symmetry axis, $\bm{\nabla}V$ is
parallel to the $z$ axis and hence $\bm{B}=0$.  Thus one easily sees
from the EOM (\ref{eq:eom_spin}) that the trajectory remains on the $z$
axis with spin frozen, and one has the frozen-spin diameter PO along
the symmetry axis.  If the potential is symmetric with
respect to the $(x,y)$ plane, one finds just as above the frozen-spin
diameter orbits in the equatorial plane with the spin parallel to the
orbital motion.
\end{enumerate}
The reduced EOM for the frozen-spin PO in the orbital plane have the same
invariance against scaling transformation (\ref{eq:scaling}),
and the action integral along the orbit is
expressed as
\begin{equation}
\oint_{\beta(e)}\bp\cdot \rmd\br=\hbar\tau_\beta\cE.
\end{equation}
Because of this simple energy dependence, equivalent to the case
without spin-orbit coupling, contributions of those POs to the
level density can be also studied conveniently with the Fourier
transformation technique.

There is another semiclassical method to
treat the spin degree of freedom, by making use of the
coupled-channel WKB formalism \cite{YabHor,LJFlyn,FriskGuhr,Amann},
where the spin is considered as a slow variable in contrast to the
orbital motion and the adiabatic approximation
is applied.  The
Hamiltonian matrix of $(2\times 2)$ spin channels is diagonalized to obtain
two adiabatic Hamiltonians, and the classical POs in those two
Hamiltonians determine the semiclassical level density.  It should be
noted that the frozen-spin POs in our approach are equivalent to
those obtained for the
\textit{diabatic} representations of Hamiltonians in the coupled-channel
WKB method \cite{FriskGuhr}.

\section{Nuclear magic numbers and pseudospin symmetry}
\label{sec:pseudospin}

Nuclear binding energies as functions of the particle
number show
remarkable fluctuation properties similar to those in the ionization
potentials of atoms.  They are both manifestation of the shell
structures for the quantized independent motion of constituent
particles in the mean fields.  In nuclear systems, quite distinct
magic numbers are known for both protons and neutrons:
\begin{equation}
N,Z=2,8,20,28,50,82,126,\cdots, \label{eq:magics}
\end{equation}
for which nuclei show the extreme stabilities.
These numbers are successfully explained by the mean-field model with
strong spin-orbit coupling, like Nilsson (modified oscillator) and WS
potential models.

An approximate dynamical symmetry called pseudospin [or pseudo SU(3)]
symmetry plays role in this shell structure
\cite{Draayer,Blokhin,BHM82}.  In the so-called pseudospin transformation,
angular momentum quantum numbers are reassigned as
$\tilde{l}=l\pm 1$ for $j=l\pm\frac12$ levels.  
The Nilsson Hamiltonian is transformed correspondingly as
\begin{align}
& H_{\rm Nils}=H_{\rm HO}-v_{ls}\bl\cdot\bs-v_{ll}\bl^2 \nonumber \\
& \to ~ 
\tilde{H}_{\rm Nils}=\tilde{H}_{\rm HO}
-(4v_{ll}-v_{ls})\tilde{\bl}\cdot\tilde{\bs}-v_{ll}\tilde{\bl}^2
-(2v_{ll}-v_{ls}).
\end{align}
Since the relation $v_{ls}\approx 4v_{ll}$ holds well, spin-orbit
coupling is quenched in the pseudospin representation, and one finds
systematic degeneracies of the pseudo spin-orbit partners
$(\tilde{j}=\tilde{l}\pm\frac12)$.
\begin{figure}
\begin{center}
\includegraphics[width=.9\linewidth]{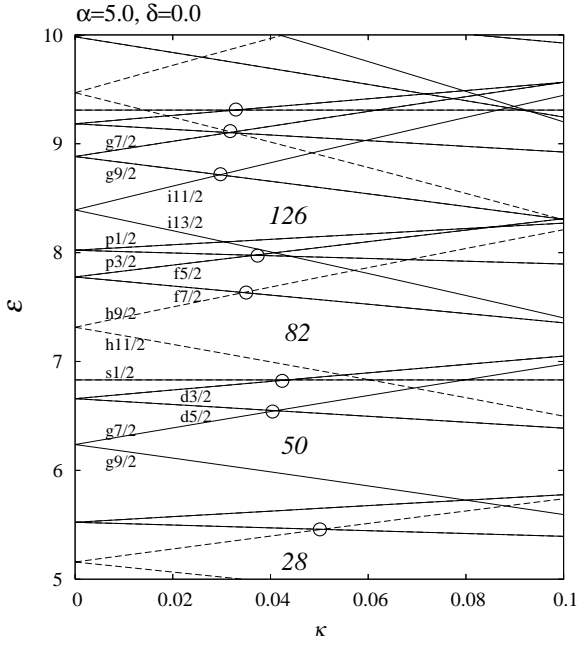}
\end{center}
\caption{\label{fig:pow_spherical}
Single-particle level diagram for the spherical power-law potential model
with the power parameter $\alpha=5.0$.  Scaled-energy levels
$\cE_j=(e_j/U_0)^{1/2+1/\alpha}$ are plotted as functions of
the spin-orbit parameter $\kappa$.  The particle numbers of the
closed-shell configurations, where all the levels below the energy gap
are occupied, are indicated in italics.  Full and broken lines
represent positive and negative parity levels, respectively.  Open dots
indicate the level crossings of the pseudo spin-orbit partners
$\tilde{j}=\tilde{l}\pm1/2$.}
\end{figure}
The same kind of level degeneracies are also found in more realistic
WS potential model, where the splittings of degenerate HO levels due
to the sharp potential surface are partially compensated by the
spin-orbit coupling.  Those quenching of the pseudo spin-orbit
splitting is considered as a result of approximate dynamical symmetry
restoration, and might be understood in relation to the
PO bifurcations as discussed in section~\ref{sec:bif}.

In the power-law potential model, surface diffuseness is controlled by
the power parameter $\alpha$, and the above development of gross shell
structure can be studied as the combinatory effect of the power
parameter $\alpha$ and the spin-orbit coupling strength $\kappa$.
Figure~\ref{fig:pow_spherical} shows the single-particle level diagram
which plots single-particle scaled energies as functions of the
spin-orbit parameter $\kappa$.  The power parameter is taken as
$\alpha=5.0$ corresponding to the medium-mass nuclei.  Systematic
degeneracies of levels are found around the realistic value of
spin-orbit strength $\kappa\approx0.05$, where a gross shell effect is
considerably developed.
Level crossings of the pseudo spin-orbit partners are indicated
by open dots.  They occur at almost the same values of $\kappa$ and
affect the gross shell structure.
The magic numbers (\ref{eq:magics}) are correctly reproduced
there.

\begin{figure}
\begin{center}
\includegraphics[width=.8\linewidth]{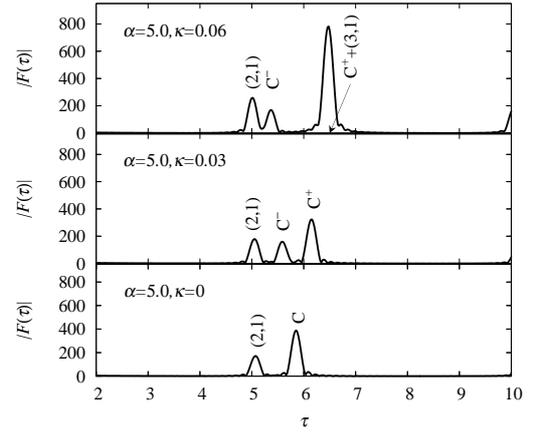}
\end{center}
\caption{\label{fig:ftk_sph}
Quantum Fourier spectra $|F^{\rm qm}(\tau)|$
calculated for the power parameter $\alpha=5.0$
with three different values of the
spin-orbit parameter: $\kappa=0$, 0.03 and 0.06.}
\end{figure}

\begin{figure}
\begin{center}
\includegraphics[width=.8\linewidth]{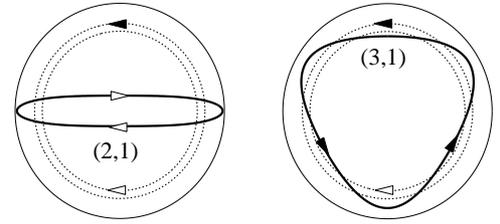}
\end{center}
\caption{\label{fig:po_sph}
Some shortest classical POs in the spherical power-law
potential model with spin-orbit coupling.  The power parameter
$\alpha=5.0$ and the spin-orbit parameter $\kappa=0.06$ are taken.  In
each panel, the outermost circle represents the boundary of the
classically accessible region, dashed lines represent the circle
orbits C$^\pm$, and thick solid lines represent the planar orbits
$(n_r,n_\phi)$.  Spin is frozen in the direction perpendicular to the
orbital plane.}
\end{figure}

As discussed in section~\ref{sec:fourier}, one can extract information
on PO contributions from the Fourier transform of
scaled-energy level density (\ref{eq:fourier_qm}).
Figure~\ref{fig:ftk_sph} shows the moduli of Fourier transform
$|F(\tau;\kappa)|$ as functions of $\tau$, for several values of the
spin-orbit parameter $\kappa$.  As expected from
equation~(\ref{eq:fourier_cl}), the Fourier amplitude shows successive
peaks at the scaled periods of classical POs
$\tau=\tau_\beta$.  For $\kappa=0$, one finds peaks at $\tau=5.1$ and
$5.8$, which correspond to the diameter orbit (2,1) and the circle
orbit C, respectively.  We label the POs by the number of
oscillation $n_r$ in radial direction and number of rotations
$n_\varphi$ about the origin, and express them as $(n_r,n_\varphi)$.
The number of radial oscillations cannot be assigned to the circle
PO and we denote it as C.  With increasing spin-orbit parameter
$\kappa$, the diameter orbit (2,1) is deformed into an oval shape, and
the circle orbit C bifurcates into C$^+$ and C$^-$ having orbital
angular momentum parallel and anti-parallel to the spin.  At
$\kappa\simeq 0.05$, the circle orbit C$^+$ undergoes bifurcation and
a new orbit (3,1) of triangular-type shape emerges.  These POs at
$\kappa=0.06$ are displayed in figure~\ref{fig:po_sph}.  As shown in
the top panel of figure~\ref{fig:ftk_sph}, the contribution of the
orbits C$^+$ and (3,1) is strongly enhanced at $\kappa=0.06$.  This is
considered as the PO bifurcation enhancement effect
which we discussed in section~\ref{sec:trace}.  Consequently, the
semiclassical origin of the development of remarkable shell structure
at $\alpha=5.0$ and $\kappa=0.05$, corresponding to medium-mass
nuclei, is shown to be related to the emergence of the orbit (3,1)
bifurcated from C$^+$ and the associated local dynamical symmetry.

\begin{figure}
\begin{center}
\includegraphics[width=\linewidth]{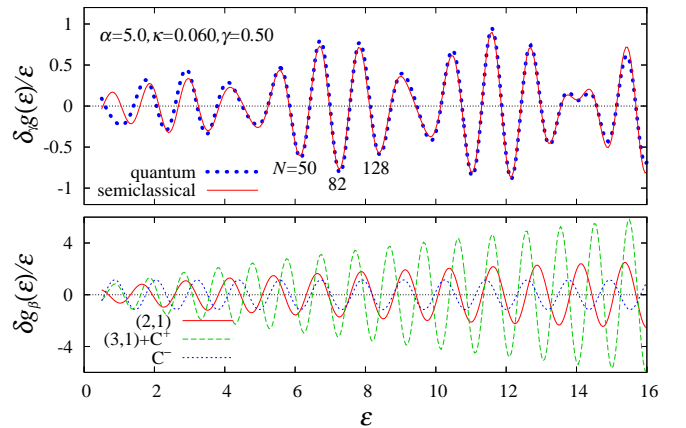}
\end{center}
\caption{\label{fig:sldfit}
Oscillating part of the scaled-energy level density for the spherical
power-law model with $\alpha=5.0$ and $\kappa=0.06$, coarse grained to
a width $\gamma=0.5$.  The upper panel compares the quantum result
and the semiclassical fit (\ref{eq:g_smooth}).  The lower panel shows
the contribution of individual POs in the semiclassical level density
(\ref{eq:dg_beta}).}
\end{figure}

The contribution of unfrozen orbit cannot be investigated by the
Fourier analysis because of the absence of the scaling, and one can
only study it by directly evaluating the semiclassical level density.
If the quantum level density is reproduced only with the contribution
of frozen-spin orbits, one may consider that the effect of unfrozen
orbits can be omitted.
Let us consider
the oscillating part of the coarse-grained scaled-energy level density
with the averaging width $\gamma$,
\begin{align}
\delta_\gamma g(\cE)
&=\frac{\gamma}{\sqrt{\pi}}\int \rmd\cE'\delta g(\cE')
 \rme^{-\left(\frac{\cE-\cE'}{\gamma}\right)^2} \nonumber \\
&=\sum_\beta \delta g_\beta(\cE)\;\rme^{-(\gamma\tau_\beta)^2/4},
\label{eq:g_smooth} \\
\delta g_\beta(\cE) &\simeq A_\beta \cE^{K_\beta/2}
 \cos(\tau_\beta\cE-\tfrac{\pi}{2}\mu_\beta). \label{eq:dg_beta}
\end{align}
The exponential damping factor $\rme^{-(\gamma\tau_\beta)^2/4}$
in (\ref{eq:g_smooth}), which appears due to the coarse-graining,
suppresses the contribution of longer POs.
Hence, the sum is dominated only by some shortest POs.  We
know the scaled
periods $\tau_\beta$ and the degeneracies $K_\beta$, but unfortunately
we have not succeeded in obtaining the semiclassical amplitudes
$A_\beta$ as well as Maslov indices $\mu_\beta$.  For the present, we
shall treat $A_\beta$ and $\mu_\beta$ as free parameters and determine
them by the least square fitting to the quantum level density.
Figure~\ref{fig:sldfit} shows the result for $\alpha=5.0$ and
$\kappa=0.06$.  We take account of the contributions of four
shortest frozen-spin POs; (2,1), C$^\pm$ and (3,1).  In the upper
panel, we compare the quantum level density with the semiclassical
fitting.  One will see that the quantum shell structure and its
beating pattern are precisely reproduced.  This seems to manifest that
the PO sum is dominated mostly by the contribution of frozen-spin POs.
In the lower panel, contributions of individual POs are shown, and
one will see that the bifurcating orbits C$^+$+(3,1) play the dominant
role in this shell structure, as indicated in the Fourier spectra
(figure~\ref{fig:ftk_sph}).  Interference with the contributions of
the other POs makes the beating pattern.  Particularly,
one sees that the
distinct magic numbers in the medium-mass region, $N=50, 82, 128$, are
established according to the constructive interference effect of those
POs.

We have also found that the above $(3,1)$ bifurcation
play significant roles in the quadrupole deformed shell structures.
It might be an interesting subject to examine
the role of the spin-orbit coupling for the properties of the
deformed shell structures and their relations to the pseudo-spin symmetry
in deformed nuclei\cite{BHM82,Dudek87,Naz90}.

\section{Bifurcations of classical periodic orbits and nuclear exotic
deformations}
\label{sec:exotic}

\begin{figure}
\begin{center}
\includegraphics[width=.6\linewidth]{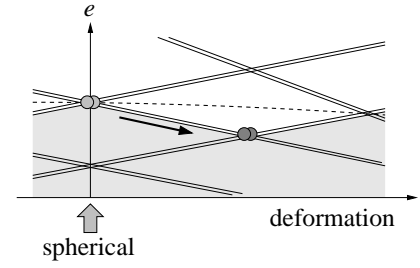}
\end{center}
\caption{\label{fig:deform}
Illustration of the mechanism of deformation induced by the spontaneous
symmetry breaking.  Solid lines represent single-particle levels and dots
denote particles in the highest partially-occupied levels.
Broken line indicate Fermi energy.
The self-bound system prefers the shape which makes level density at
Fermi surface as low as possible to make the largest shell energy gain.}
\end{figure}

In the classical regime, self-bound interacting many-body system
favors the spherical shape, since the system prefers the shape whose
surface area is as smaller as possible under the fixed volume.  In the
quantum regime, the quantum shell effects evoke various deformations
to the system.  These shell effects are caused by the fluctuation in
the single-particle spectra.  Nuclei show particular stability at the
spherical shape when the levels under the energy gap are completely
occupied.  The magic numbers (\ref{eq:magics}) correspond to such
closed-shell configurations.  In situation where the degenerate levels
at the Fermi energy are partially occupied, system tends to deform in
order to lower the energy by splitting the degenerate levels by
deformation, as illustrated in figure~\ref{fig:deform}.
The way of the level splittings depends on the types of the deformations,
and such shape that makes the level density at the Fermi energy lower
is preferred.  It is a kind of spontaneous breaking of symmetry
similar to the Jahn-Teller effect known in the molecular systems.  In
this section, we discuss some nuclear exotic deformations and their
semiclassical origins with the use of the POT.

\subsection{Superdeformations}
\label{sec:sd}

Concerning nuclear deformations, one of the most exciting discovery is
the so-called \textit{superdeformed states} in rapidly rotating
nuclei, having extremely large quadrupole deformation whose axis ratio
amounts to 2:1 \cite{NT88,JK91}.
Nuclei with such large deformations are
also found in the fission process as isomers formed between
the double-humped potential barriers\cite{FunnyHills,BL80}.
The search of the second minima having much larger deformation whose axis
ratio close to 3:1, often referred to as \textit{hyperdeformed states}, is
also a hot subject in the high-spin nuclear physics for both theories
and experiments\cite{HYP}.
For such large deformations to be realized, significant shell energy gain
should be provided particularly at those shapes
in addition to the macroscopic
driving force like Coulomb repulsion and rotation.

\begin{figure}
\begin{center}
\includegraphics[width=\linewidth]{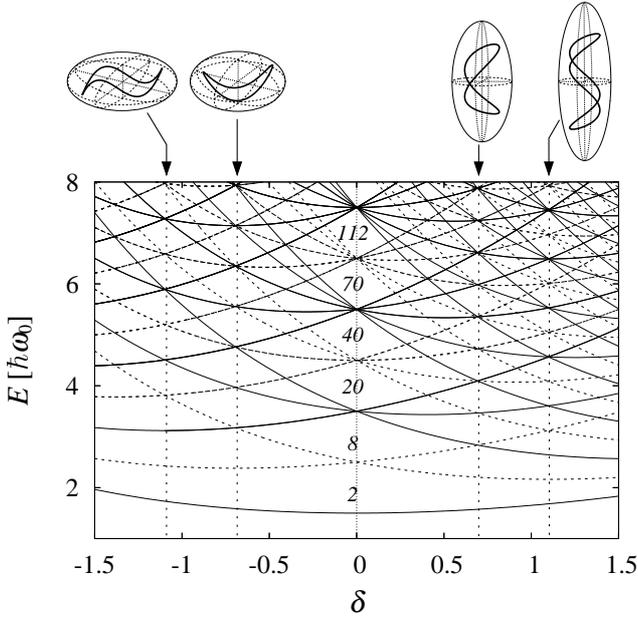}
\end{center}
\caption{\label{fig:nils_ho}
Single-particle level diagram for the axially deformed harmonic
oscillator model.  Energy eigenvalues are plotted as functions of
deformation parameter $\delta$.  The particle numbers of the spherical
closed-shell configurations, with the spin degeneracy factor taken
into account, are indicated in italics.  On the top of
the diagram, typical classical POs at prolate and oblate super and
hyper-deformed shapes are shown.}
\end{figure}

In the following, let us investigate the emergence of the
superdeformed shell structures and
their semiclassical origins.
Here, we neglect the spin-orbit
coupling for simplicity.  The simplest model for describing the
superdeformed shell structure is the axially-symmetric harmonic
oscillator (HO)
\begin{equation}
H_{\rm HO}=\frac{\bp^2}{2m}+\frac{m\{\omega_\perp^2(x^2+y^2)+\omega_z^2
z^2\}}{2} \label{eq:h_ho}
\end{equation}
with volume conservation condition $\omega_\perp^2\omega_z=\omega_0^3$.
Energy eigenvalues are given analytically by
\begin{equation}
e_{n_\perp n_z}=\hbar\omega_\perp(n_\perp+1)
+\hbar\omega_z(n_z+\tfrac12),
\end{equation}
and the simultaneous degeneracies of levels take places where the
frequencies $\omega_\perp$
and $\omega_z$ become commensurable.
Figure~\ref{fig:nils_ho} displays the single-particle level diagram,
in which the energy eigenvalues are plotted as functions of the
deformation
parameter $\delta=\log(\omega_\perp/\omega_z)$.
Particularly, one sees prominent shell structures at
$\rme^\delta=2^{\pm 1}$ $(\delta=\pm 0.693)$ and $\rme^\delta=3^{\pm
1}$ $(\delta=\pm 1.099)$, which correspond to the superdeformed and
hyperdeformed shapes, respectively.

In semiclassical POT, those shell structures are
understood as the result of the emergence of the four-parametric
PO families at the deformations with rational frequency
ratios.  Typical POs at those deformations are shown in
the top of figure~\ref{fig:nils_ho}.  In the HO model, these
degenerate PO families can exist only at the deformations
with rational axis ratios.

\begin{figure}
\begin{center}
\includegraphics[width=\linewidth]{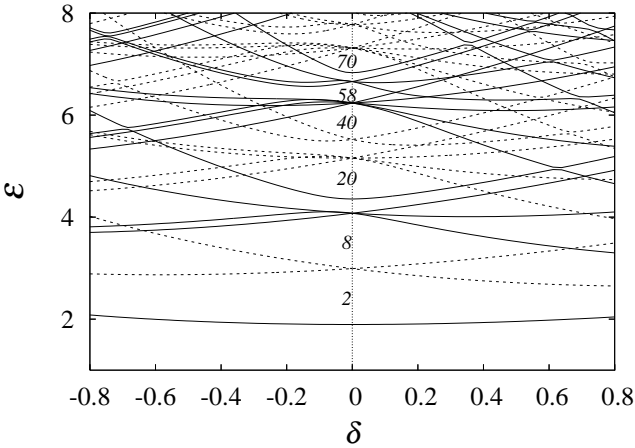}
\end{center}
\caption{\label{fig:nils_a50}
Single-particle level diagram for the spheroidal power-law potential
model with the power parameter $\alpha=5.0$.  Scaled-energy levels
$\cE_j=(e_j/U_0)^{1/2+1/\alpha}$ are plotted as functions of the
deformation parameter $\delta$.  The particle numbers of the
closed-shell configurations at the spherical shape are indicated
in italics.}
\end{figure}

On the other hand, realistic nuclear mean field potential has
sharper surface with increasing mass number.  Let us
consider the deformed shell structure in the radial power-law
potential model with spheroidal deformation, where the shape function
$f$ in equation~(\ref{eq:h_power}) is given by
\begin{equation}
f(\theta;\delta)=\frac{1}{\sqrt{\rme^{-\frac43\delta}\cos^2\theta
 +\rme^{\frac23\delta}\sin^2\theta}}.
\end{equation}
This model is integrable in the two limits: $\alpha=2$
(axially deformed HO) and $\alpha=\infty$ (spheroidal cavity),
and nearly integrable between them: a large portion of the classical
phase space is foliated with the KAM tori.
Figure~\ref{fig:nils_a50} shows the single-particle level diagram of
the power-law potential model with the power parameter $\alpha=5.0$.  One
finds level bunchings around the superdeformed region $|\delta|\sim 0.7$
although they are less clear compared with the case of HO.

Let us analyze the properties of classical POs in the
spheroidal power-law potential to investigate the classical-quantum
correspondence.
In the HO limit, $\alpha=2$, all the classical motions are periodic at
the spherical shape ($\delta=0$).  Varying $\alpha$ from the HO value,
only the circle and diameter POs survive.  If the
potential is deformed into spheroidal shape, the circle PO family
bifurcate into the meridian oval family C and the isolated equatorial
circle EC.  The diameter PO also bifurcate into the degenerate
equatorial diameter family X and the isolated symmetry-axis diameter Z.
\begin{figure}
\begin{center}
\includegraphics[width=\linewidth]{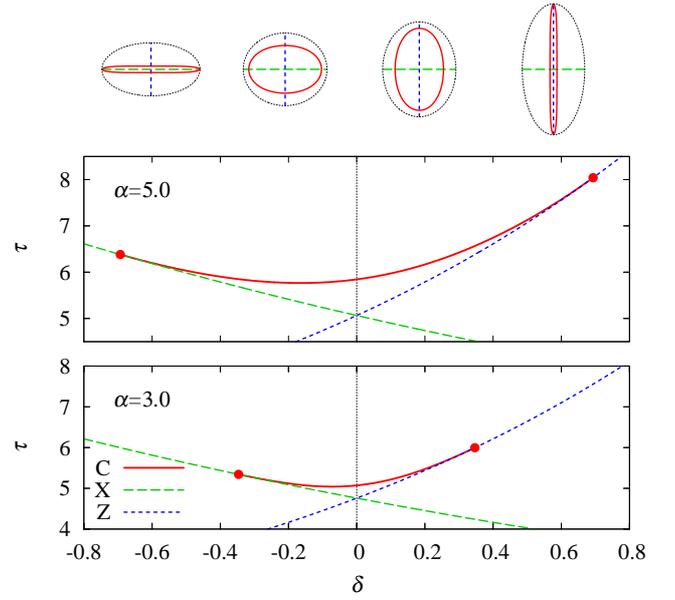}
\end{center}
\caption{\label{fig:bridge_nd}
Scaled periods $\tau_\beta$ of the orbits C, X and Z plotted as
functions of the spheroidal deformation parameter $\delta$ for the
power parameter $\alpha=3.0$ (lower panel) and 5.0 (upper panel).  The
orbit C makes a bridge between the orbits X and Z.  Solid dots
indicate the bifurcation points.  Those three POs for $\alpha=5.0$ at
several values of $\delta$ are displayed on the top.}
\end{figure}
Figure~\ref{fig:bridge_nd} shows the scaled periods of those classical
POs for several values of $\alpha$ as functions of the
deformation parameter $\delta$.  With increasing prolate deformation
$\delta>0$, the meridian oval orbit C is continuously deformed and
finally submerge into symmetry-axis diameter Z at certain deformation
$\delta_c$.  With increasing oblate deformation $\delta<0$, the orbit
C is deformed in a different way and finally submerge into the
equatorial diameter X at the deformation $-\delta_c$.  In this way,
the oval orbit family C make a bridge between the two bifurcations
from the diameter orbits X and Z with varying deformation.  We call
such kind of bifurcation scenario as the
\textit{bridge orbit bifurcation}.  The classical and semiclassical
analyses of the bridge orbit bifurcations are given in
\cite{AriBr2008b} with various practical examples.  It should be
emphasized that the two orbits connected by the bridge are widely
separated from each other in the phase space.  If the bridge is short
enough in the deformation space, the dynamical symmetry restored at
one end of the bridge will be approximately kept along the bridge to
the other end, and it may bring about a family of quasi-periodic
orbits occupying much larger phase space volume than that
in case of simple bifurcations.

\begin{figure}
\begin{center}
\includegraphics[width=.9\linewidth]{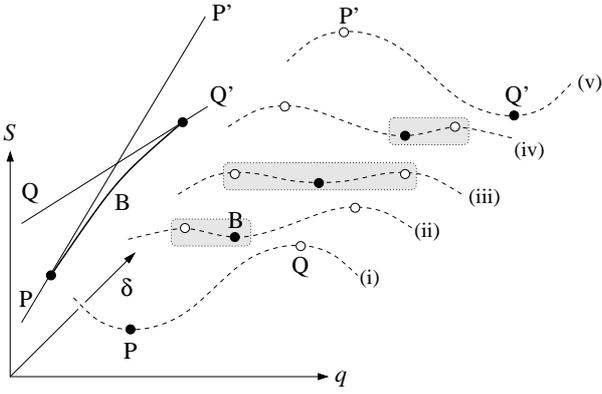}
\end{center}
\caption{\label{fig:bif_bridge}
Illustration of the bridge-orbit bifurcation scenario.  Broken lines
represent $S(q)$, the action integral along the closed orbit starts
from $q$ and returns to the same point $q$, at several deformations in
the bifurcation process.  Stationary points of $S(q)$ give the POs.
With varying deformation $\delta$, a bridge orbit B emerges from the
orbit P and then submerge into the orbit Q.  A family of
quasi-periodic orbits is formed around the shaded area.  Full lines
drawn in the $(\delta,S)$ plane are the action integrals along the
POs, and dots indicate the bifurcation points.
Reproduced with permission from \cite{AriMuk}.
Copyright American Physical Society 2014.}
\end{figure}

Figure~\ref{fig:bif_bridge} illustrates the scenario of the
bridge-orbit bifurcation:
\begin{enumerate} \def\labelenumi{(\roman{enumi})}\itemsep=0pt
\item There are two different POs, P and Q, corresponding
to the two stationary points of the action function $S(q)$, which are
widely separated from each other.
\item With increasing deformation $\delta$, the orbit P undergoes bifurcation
and a new orbit B emerges from it.  One finds a family of quasi-periodic
orbits around these stationary points.
\item Action integrals of P and Q orbits crosses in the $(\delta,S)$ plane,
and the quasi-periodic family extends from P to Q, implying a development
of large dynamical symmetry around them.
\item The orbit B approaches the orbit Q and
\item finally submerges into the orbit Q.
\end{enumerate}

In comparison to the simple bifurcations which may cause
local dynamical symmetries only in vicinity of the single bifurcating
PO, we could expect the bridge orbit to give
much more significant effect on the quantum shell effect due to the
large phase-space volume of the quasi-periodic orbit family formed
around the bridge orbits.  From figure~\ref{fig:bridge_nd}, one will
note that existence domain $(-\delta_c,\delta_c)$ for the bridge
orbit C grows as the power parameter $\alpha$ becomes
larger.  The contribution of bridge orbit might be less important as
the above domain grows due to the breaking of the dynamical
symmetry between the two ends of the bridge, and thus, shell effect
is generally reduced as $\alpha$ increases.

\begin{figure}
\begin{center}
\includegraphics[width=\linewidth]{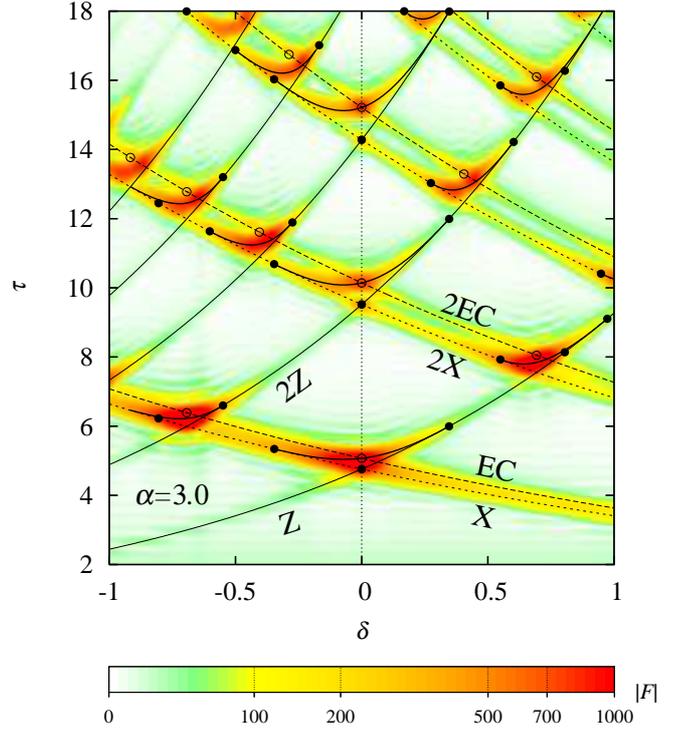}
\end{center}
\caption{\label{fig:ftl_nd}
Color map of the quantum Fourier amplitude
$|F^{\rm qm}(\tau;\delta)|$ in the $(\delta,\tau)$ plane.  The power
parameter $\alpha=3.0$ is taken.  Lines represent the scaled periods
$\tau_\beta(\delta)$ of some classical POs as functions of
$\delta$.  Solid and open dots indicate the bifurcation points of
the meridian orbits and equatorial orbits, respectively.  EC, X and Z
represent the equatorial circle, equatorial diameter and symmetry-axis
orbits, respectively, and 2EC, 2X, 2Z are their second repetitions.}
\end{figure}

To investigate the contribution of these orbits to the level
density, we calculate the Fourier transforms of the scaled-energy level
density (\ref{eq:fourier_qm}).  Figure~\ref{fig:ftl_nd} shows the
Fourier amplitudes $|F(\tau;\delta)|$ plotted in the $(\delta,\tau)$
plane.  The power parameter $\alpha=3.0$, a little larger
than the HO
value, is taken as an illustration.  The scaled period
$\tau_\beta(\delta)$ of the classical POs are also drawn
in the same plane.  One finds an excellent correspondence between Fourier
peaks and the classical POs.  The Fourier amplitudes take
especially large values along the bridge orbits appearing at each
crossings of the repetitions of the equatorial and symmetry axis orbits.
One may also note that the Fourier amplitudes along the orbit $n$X
($n$\,th repetition of X) are
larger than those along the orbit $n$Z.  This is because the orbit X
forms a one-parametric family with respect to the rotation about the
symmetry axis, while the orbit Z is isolated.  In the
superdeformed region $\delta\approx 0.6$, the bridge orbits between the
second repetition of equatorial orbits and the primitive symmetry-axis
orbit play important role.

\begin{figure}
\begin{center}
\includegraphics[width=\linewidth]{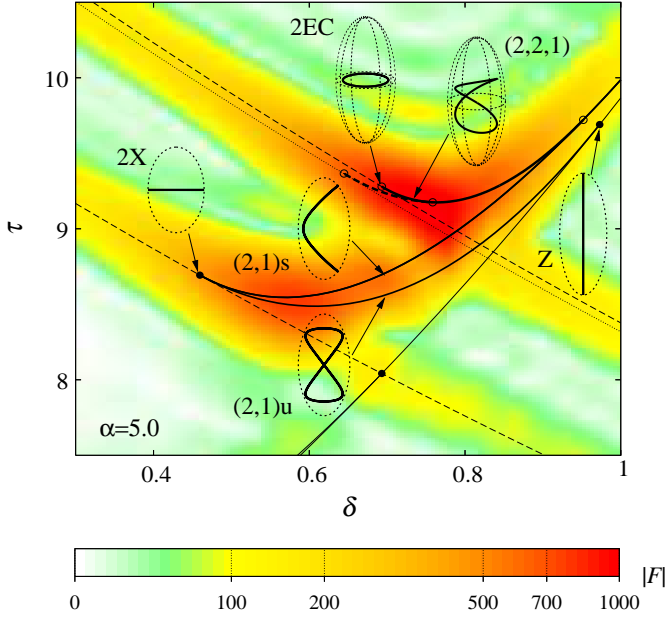}
\end{center}
\caption{\label{fig:ftl_sd}
Color map of the quantum Fourier amplitude $|F^{\rm qm}(\tau,\delta)|$ for
$\alpha=5.0$ in the superdeformed region.
Lines represent the scaled periods $\tau_\beta(\delta)$ of the classical
POs as functions of $\delta$, and dots indicate the bifurcation
points.  In each inserted figure, the PO is drawn with
thick solid line, and the boundary of the classically accessible region
is indicated by dotted ellipse(s).}
\end{figure}

In figure~\ref{fig:ftl_sd}, we examine the Fourier spectra in
superdeformed region in detail taking the power parameter $\alpha=5.0$
suitable for medium-mass nuclei.  In the superdeformed region,
the equatorial diameter X undergoes period-doubling bifurcation at
$\delta=0.46$ and a pair of stable and unstable meridian orbits (2,1)s
and (2,1)u emerge.  Here, the meridian orbits in the $(x,z)$ plane
are labeled by the numbers of oscillations in the $x$ and $z$ directions
$(n_x,n_z)$.  The above meridian orbits
change their shapes with increasing $\delta$, and
finally, (2,1)u submerge into the symmetry-axis orbit Z at $\delta=0.97$,
and (2,1)s submerge into Z at $\delta=1.28$.  Namely, there are two
bridge orbits between equatorial diameter and symmetry-axis orbit.
One also sees another bridge orbit between the second repetition of
equatorial circle orbit 2EC and symmetry-axis orbit Z around a little
larger deformation.  The orbit EC undergoes period-doubling
bifurcation at $\delta=0.69$ and a new three-dimensional
(3D) orbit (2,2,1) emerges.  3D orbits are
labeled by the numbers of
oscillations (rotations) $(n_\rho,n_\varphi,n_z)$ in the directions of
the cylindrical coordinate ($\rho$, $\varphi$, $z$).
With increasing
$\delta$, the orbit (2,2,1) first submerge into meridian orbit (2,1)s
(the stable branch of the meridian bridges) at $\delta=0.95$ before
finally submerge into the orbit Z at $\delta=1.28$.  One sees Fourier
amplitude greatly enhanced along these bridge orbits and they should
play most significant roles in emergence of superdeformed shell
structure.

\begin{figure}
\begin{center}
\includegraphics[width=.9\linewidth]{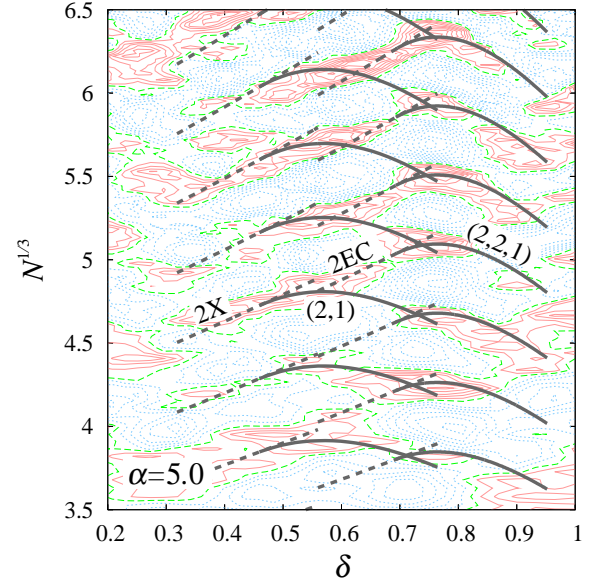}
\end{center}
\caption{\label{fig:sce_sd}
Contour map of the shell energy $\delta E(N;\delta)$ plotted in the
$(\delta,N^{1/3})$ plane.  Solid (red) and dashed (blue) contour lines
represent negative and positive $\delta E$, respectively.  Thick
lines represent the constant-action ones (\ref{eq:cac}) for some
short classical POs in the portion where they make
dominant contributions to the trace formula.}
\end{figure}

Next, let us evaluate the shell energies $\delta E(N)$ as functions of
deformation, and examine the effect of above bifurcations.
Suppose the situation where a single orbit $\beta$ dominates the PO sum
in equation~(\ref{eq:trace_esh}), namely,
\begin{gather}
\delta E(N) \approx \left(\frac{\hbar}{T_\beta(e_F)}\right)^2
A_\beta(e_F)\cos\left(\frac{1}{\hbar}S_\beta(e_F)
 -\frac{\pi}{2}\mu_\beta\right).
\end{gather}
Then, the shell energy takes minima where the conditions
\begin{equation}
\tau_\beta(\delta)\cE_F(N)-\frac{\pi}{2}\mu_\beta
=(2n+1)\pi, \quad n=0,1,2,\cdots
\end{equation}
are satisfied.  Using the Thomas-Fermi approximation (\ref{eq:geps_TF}),
Fermi energy $\cE_F$ is approximately given as
\begin{subequations}
\begin{gather}
N\approx\int_0^{\cE_F}g_{\rm TF}(\cE)\rmd\cE
=\frac{c_0}{3}\cE_F^3, \\
\cE_F\approx\left(\frac{3N}{c_0}\right)^{1/3}.
\end{gather}
\end{subequations}
Therefore, the shell energy will present valleys
along the \textit{constant-action lines} \cite{StrMOD}
\begin{equation}
N^{1/3}=\left(\frac{c_0}{3}\right)^{1/3}
\frac{(2n+1+\mu_\beta/2)\pi}{\tau_\beta(\delta)}\,,
\quad n=0,1,2,\cdots
\label{eq:cac}
\end{equation}
in the $(\delta,N^{1/3})$ plane.  Figure~\ref{fig:sce_sd} shows contour
map of the shell energy $\delta E(N;\delta)$ in the $(\delta,N^{1/3})$
plane.  One sees regular and strong oscillations in $\delta E(N)$
to develop around $\delta\approx 0.6$, which is considered as the
effect of the superdeformed
shell structure.  Thick curves represent the constant-action lines
(\ref{eq:cac}) of some short classical POs.  Shell energy valleys in
the region $\delta=0.4\sim 0.6$ are nicely explained by the meridian
(2,1) bridge orbits, and those in the region $\delta=0.7\sim 0.9$ are
by the 3D orbit (2,2,1), just as expected from Fourier
analysis.  Hence, we can conclude that the bridge orbit bifurcations
between the second repetition of equatorial and
the primitive symmetry-axis orbits are
responsible for the emergence of superdeformed shell structure.  This
is a general consequence
valid for any value of $\alpha$ from HO to
cavity values\cite{Magner2002}, any other parametrization of
quadrupole shapes, with and without spin-orbit
coupling\cite{Arita2004}.

\subsection{Octupole deformations}
\label{sec:octupole}

The effect of reflection-asymmetric octupole degrees of freedom is
also an important issue in nuclear structure physics\cite{BN1996}.
Most of the nuclei are known to have reflection-symmetric ground
states, and the violation of this fundamental symmetry may provide us
with valuable information on the nuclear dynamics.  As reviewed in
\cite{BN1996}, several static octupole-deformed states have been
observed, e.g., through the low-lying negative parity states and the
parity-doublet rotational bands connected with E1
and E3 transitions.  It is also predicted that the excited rotational
states have quite unique nature when they are build on the ground
state having an octupole shape with the point group symmetry such as
the tetrahedral one\cite{Tagami2013}.  Since no driving
forces towards reflection-asymmetric shapes are found in the
classical dynamics, quantum shell effects are considered as the
exclusive origin of the octupole deformations.

The reflection asymmetries are
also important in description of the asymmetric fission processes
of heavy elements\cite{FunnyHills,GMN71}.  The semiclassical
POT is also useful in accounting for the formation of fission path
towards the reflection-asymmetric shapes\cite{BRS97}.

Hamamoto \etal have investigated the octupole deformed shell
structures by considering four kinds of pure octupole deformations
added to the spherical potential\cite{HMXZ}.  They found a remarkable
shell structure develops at finite $Y_{32}$-type deformation which has
the tetrahedral $T_d$ symmetry.  The importance of the tetrahedral
deformation is also discussed for nuclei\cite{Takami,Dudek2002} 
and metallic clusters\cite{Reimann}.  Here we are going to
extend the analysis of \cite{HMXZ} to a more realistic power-law
potential model and investigate the semiclassical origins of
octupole-deformed shell structures.  There are several ways of
parametrizing octupole shapes.  In the WS model, the shape of the
equi-potential surface is usually parametrized as
\begin{subequations}\label{eq:oct_par1}
\begin{align}
r=& R_0(1+\beta_{3m}\tilde{Y}_{3m}), \\
& \tilde{Y}_{3m}=\sqrt{2-\delta_{m0}}\Re Y_{3m}.
\end{align}
\end{subequations}
In the modified oscillator model employed by \cite{HMXZ}, octupole
potential is introduced in addition to the spherical central potential
as
\begin{equation}
V(\br)=\frac{m\omega_0^2r^2}{2}\left[1-2\beta_{3m}\tilde{Y}_{3m}\right].
\end{equation}
In this case, the shape of the equi-potential surface is expressed as
\begin{equation}
r=R_0[1-2\beta_{3m}\tilde{Y}_{3m}]^{-1/2}. \label{eq:oct_par2}
\end{equation}
Above parametrizations can be generalized to a formula
\begin{equation}
r=R_0[1+k\beta_{3m}\tilde{Y}_{3m}]^{1/k} \label{eq:oct_pargen}
\end{equation}
which corresponds to (\ref{eq:oct_par1}) for $k=1$ and to
(\ref{eq:oct_par2}) for $k=-2$, respectively.  This generalized
formula gives the identical shape independent of $k$ up to the first
order of $\beta_{3m}$, while it gives considerably different shapes
dependent on $k$ for large $\beta_{3m}$.  To obtain the optimum shape
parametrization, we consider minimization of the area of the
equi-potential surface with respect to $k$ under the fixed volume
surrounded by the surface.  For a given $\beta_{3m}$ with varying $k$,
the surface area is found to take minimum around $k=0$.  Hence we take
the $k\to0$ limit of equation~(\ref{eq:oct_pargen}), which results in
an exponential function.  Then, our Hamiltonian is expressed as
\begin{equation}
H=\frac{p^2}{2M}+U\left[\frac{r}{R_0(\beta_{3m})
 \exp(\beta_{3m}\tilde{Y}_{3m})}\right]^\alpha.
\label{eq:shape_oct}
\end{equation}
$R_0(\beta_{3m})$ is determined by the volume
conservation condition.

\begin{figure}
\begin{center}
\includegraphics[width=.85\linewidth]{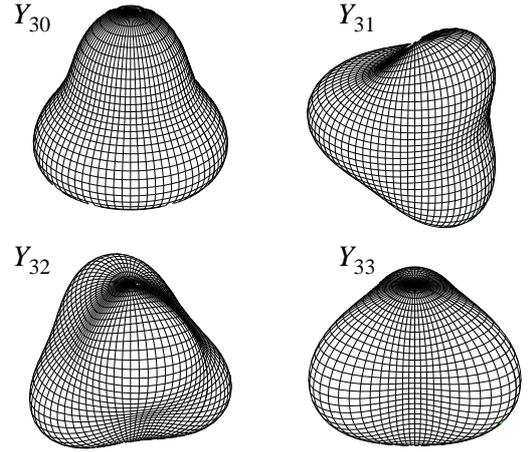}
\end{center}
\caption{\label{fig:shape_oct}
Equi-potential surfaces for octupole deformed potentials
(\ref{eq:shape_oct}) for $\beta_{3m}=0.4$.}
\end{figure}

Figure~\ref{fig:shape_oct} displays the equi-potential surfaces for
four types of purely octupole-deformed potentials at the octupole
parameters $\beta_{3m}=0.4$.  $Y_{30}$ shape has a continuous axial
symmetry, while the other shapes have different kinds of discrete
point-group symmetries\cite{LLQM,GroupTxt}.  Those symmetries
can be
utilized in quantum calculations to classify the eigenstates according
to the irreps (irreducible representations) of the symmetry group.
The $Y_{31}$ and $Y_{33}$ shapes have $C_{2v}$ and $D_{3h}$
symmetries, respectively, which have up to two-dimensional irreps.
The $Y_{32}$ shape has the tetrahedral ($T_d$) symmetry consists of 24
different symmetric transformations and has
three-dimensional irreps.
Since the degeneracy factor of the levels is equal to the dimension of
the irrep, one generally find levels with three-fold degeneracies in
the $Y_{32}$-deformed states.  Due to this higher degeneracies, the
$Y_{32}$ deformed states are expected to have stronger shell effect
than those with the other types of octupole shapes.

In addition to the above geometrical degeneracy effect, Hamamoto \etal
have found a strong bunching of levels for finite $Y_{32}$
deformation, and the shell-energy gains with $Y_{32}$ deformation
may surpass those with quadrupole deformations in certain particle
number regions.

\begin{figure}
\begin{center}
\includegraphics[width=\linewidth]{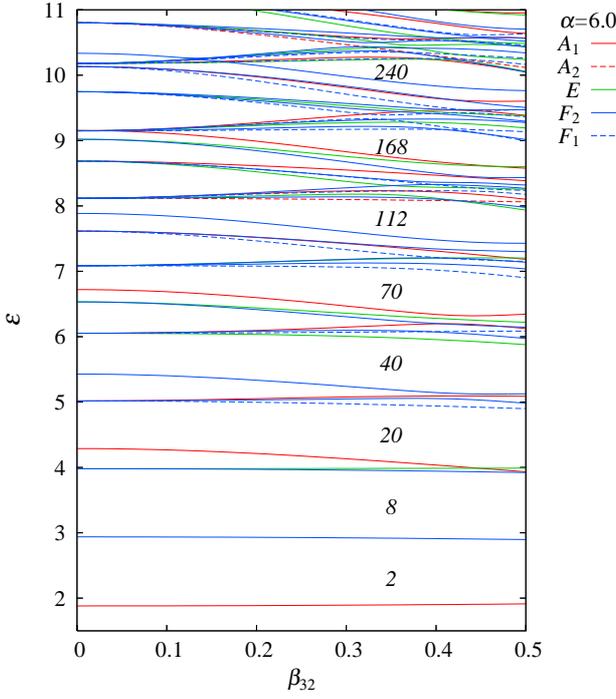}
\caption{\label{fig:sps_y32}
Single-particle level diagram of the octupole-deformed power-law
potential model with the power parameter $\alpha=6.0$.  Scaled-energy
eigenvalues are plotted as functions of $Y_{32}$ deformation parameter
$\beta_{32}$.  Red solid and dashed lines represent the
one-dimensional
irreps $A_1$ and $A_2$, respectively, which have no degeneracies.
Green solid lines represent the
two-dimensional irrep $E$ which are
doubly degenerate.  Blue solid and dashed lines represent the
three-dimensional irreps $F_2$ and $F_1$, respectively, which are
triply degenerate.  The particle numbers of the closed-shell
configurations around $\beta_{32}=0.3\sim 0.4$ are indicated in italics.}
\end{center}
\end{figure}

Figure~\ref{fig:sps_y32} shows the single-particle level diagram with
the power parameter $\alpha=6.0$, where the single-particle scaled
energies $\cE_j$ are plotted as functions of $Y_{32}$-deformation
parameter $\beta_{32}$.  The degenerate levels at the spherical shape
split with increasing octupole deformation, but they eventually form a
pronounced shell structure around $\beta_{32}=0.4$.  Surprisingly,
the particle numbers corresponding to the closed-shell configurations are
equivalent to those of spherical HO model\cite{Reimann}.  Although the
obtained shell effects here are not as strong as what we have obtained
in \cite{AriMuk} using the shape parametrization interpolating sphere and
tetrahedron, qualitative features are quite similar.  One finds no
such remarkable shell structures for the other types of octupole shapes.

\begin{figure}
\begin{center}
\includegraphics[width=\linewidth]{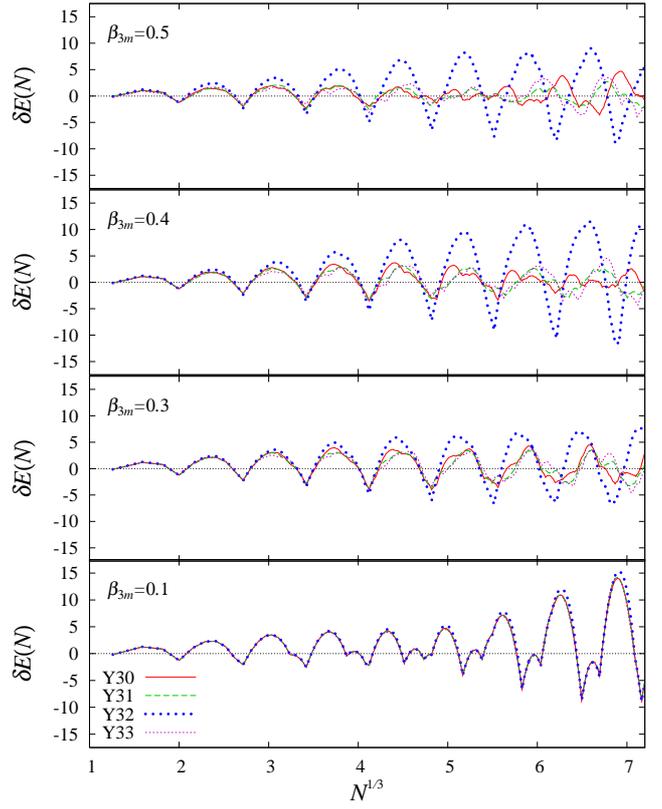}
\end{center}
\caption{\label{fig:shell_oct}
Shell energies as function of particle number $N$ for several octupole
parameters $\beta_{3m}$ with the power parameter $\alpha=6.0$.
Red (solid), green (dashed), blue (thick dotted) and magenta (thin
dotted) lines represent $Y_{30}$, $Y_{31}$,
$Y_{32}$ and $Y_{33}$ deformations, respectively.}
\end{figure}

In figure~\ref{fig:shell_oct}, we compare deformed shell energies for
different types of octupole shapes, taking the power parameter
$\alpha=6.0$.  Shell energies $\delta E(N)$ are plotted as functions
of the particle number $N$ for several values of deformation
parameters.  For small $\beta_{3m}$, shell energies show supershell
structures due to the interference of two groups of the POs: ones
bifurcated from the circle orbit and the others from the
diameter orbit.  As the octupole deformation parameter increase, the
fluctuations in shell energies show different structures and
amplitudes for different types of
deformation.
One finds that the gross shell effects are remarkably
enhanced for the
$Y_{32}$ deformation:  It shows quite
regular oscillations and are most developed around $\beta_{32}=0.4$,
as expected from the level diagram shown in figure~\ref{fig:sps_y32}.

\begin{figure}
\begin{center}
\includegraphics[width=.8\linewidth]{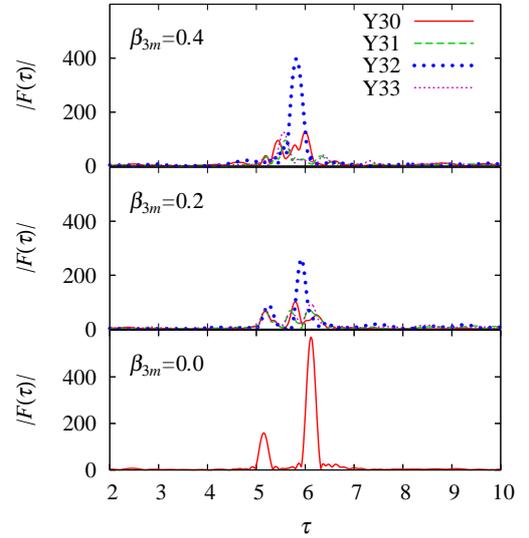}
\end{center}
\caption{\label{fig:fourier_oct}
Quantum Fourier spectra $|F^{\rm qm}(\tau;\beta_{\rm 3m})|$
for the power-law potential models of spherical and various
octupole shapes.  The bottom, middle and top panels are for
$\beta_{3m}=0.0$ (spherical), 0.2 and 0.4, respectively.  The results
for the four different types of octupole shapes
$\tilde{Y}_{30}\sim\tilde{Y}_{33}$
are plotted with red (solid), green (dashed), blue (thick dotted) and
magenta (thin dotted) lines, respectively.}
\end{figure}

Figure~\ref{fig:fourier_oct} shows the Fourier spectra $|F(\tau)|$
calculated for the octupole deformation parameters $\beta_{3m}=0$, 0.2
and 0.4.  For the spherical shape $\beta_{3m}=0$, one sees two
prominent peaks at $\tau=5.15$ and $6.11$ corresponding to the diameter
and circle POs, respectively.  These peaks rapidly decreases with
increasing $\beta_{3m}$ for $m\ne 2$, while the peak at $\tau\sim
6.0$ remains large for $Y_{32}$ deformation.  At $\beta_{32}\sim 0.4$,
this peak is enhanced again and one might expect that the
corresponding PO will make significant contribution to
the level density at this deformation.

\begin{figure}
\begin{center}
\includegraphics[width=\linewidth]{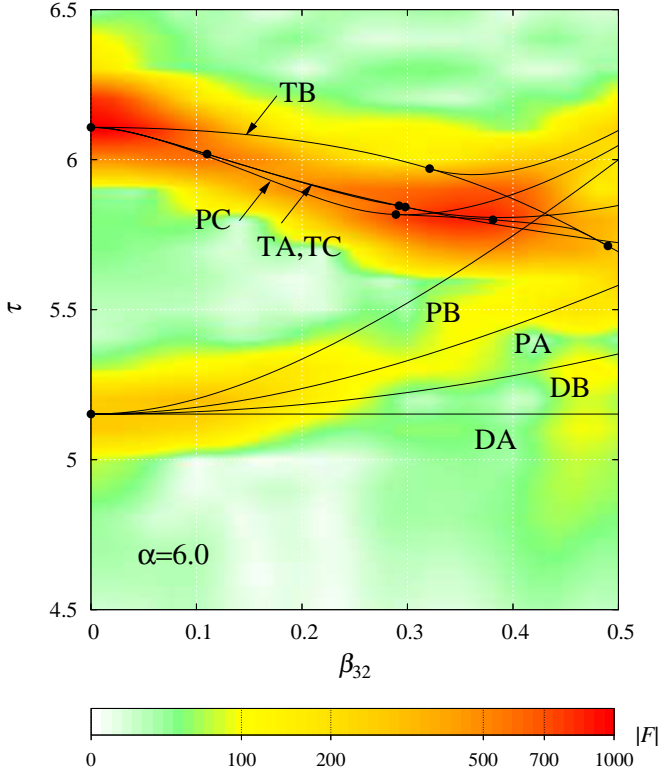}
\end{center}
\caption{\label{fig:ftl_oct}
Color map of the quantum Fourier amplitude $|F^{\rm qm}(\tau;\beta_{32})|$ for
$\alpha=6.0$ plotted in the $(\beta_{32},\tau)$ plane.  Curves represent
scaled periods of the classical POs plotted
as functions of octupole parameter $\beta_{32}$, and dots indicate
their bifurcation points.}
\end{figure}

\begin{figure}
\begin{center}
\includegraphics[width=\linewidth]{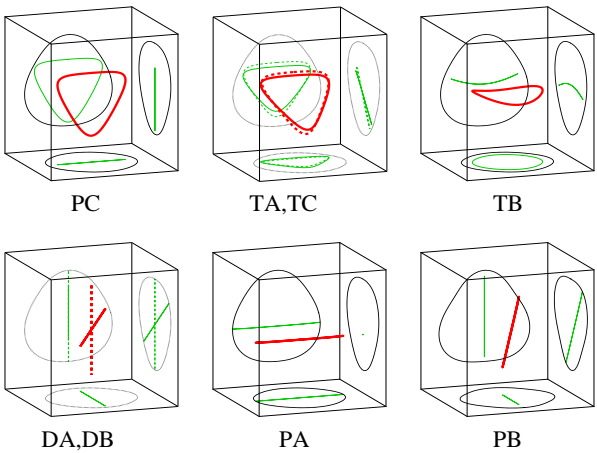}
\end{center}
\caption{\label{fig:po_y32}
Some shortest classical periodic orbits for $Y_{32}$ deformed state
at the octupole parameter $\beta_{32}=0.2$ with the power parameter
$\alpha=6.0$.  Their projections onto the
$(x,y)$, $(y,z)$ and $(z,x)$ planes are also shown, with the
boundaries of the classically accessible region.}
\end{figure}

To elucidate the origin of the emergence of this remarkable shell
structure associated with $Y_{32}$ deformation, we examine in
figure~\ref{fig:ftl_oct} the correspondence between the distribution
of the Fourier peaks and the scaled-periods of the classical POs in
the $(\beta_{32},\tau)$ plane.  Some shortest POs are displayed in
figure~\ref{fig:po_y32} for $\beta_{32}=0.2$.  Here, we name each
PO with two characters: The first character ``D'', ``P'' or ``T''
stands for diameter, planar or three-dimensional; the second one
is put alphabetically in the order we've found them.  With increasing
$\beta_{32}$, the diameter orbit bifurcates into four different
orbits: the diameter DA along the three-fold rotation axis, the diameter DB
along the four-fold rotatory reflection axis, librational orbits PA
and PB in the mirror-symmetry plane.  From the Fourier analysis, their
contributions to the level density are monotonically reduced with
increasing $\beta_{32}$.  The circle orbit bifurcates into three orbits:
the isosceles triangular-type orbit PC in the mirror plane, the
equilateral triangular-type orbit TA having three-fold rotational
symmetry, and the square-type orbit TB having four-fold rotatory
reflection symmetry.  The orbit PC undergoes bifurcation at
$\beta_{32}=0.035$ from which a 3D orbit TC emerges.
The orbits TA and TC undergo so-called {touch-and-go bifurcation} at
$\beta_{32}=0.11$.  As a common property in these three orbits, the
monodromy matrix has an eigenvalue which is kept close to unity up to
large values of $\beta_{32}$.  As we discussed in
section~\ref{sec:bif} a local family of quasi-periodic orbit is formed
around such an orbit and it makes coherent contribution to the trace
integral.  This explains the reason why the contribution of these
orbits remain large with increasing $\beta_{32}$.  They undergo
bifurcations almost simultaneously around $\beta_{32}\sim 0.3$ and
yields new POs, some of which make bridges between them.
As we see in figures~\ref{fig:fourier_oct} and \ref{fig:ftl_oct},
significant enhancement of
the Fourier peak corresponding to those bifurcations is found.  Some
details on these bifurcations are described in the appendix.  Since
these orbits have almost the same values of scaled periods
$\tau_\beta$, they bring about a quite regular shell structure.  The
approximate coincidence of their actions and the almost simultaneous
occurrence of bifurcations generating the bridge-orbit networks
connecting them strongly suggest the underlying dynamical symmetry.
This symmetry restoration, caused almost simultaneously around many
different POs and also mapped onto their replicas generated by the 24
symmetry transformations of $T_d$, is considered to develop into in somewhat
global one.  Recalling the magic numbers at $\beta_{32}\approx 0.4$
shown in figure~\ref{fig:sps_y32} which are equivalent to those of
spherical HO, one may surmise that a restoration of the dynamical
symmetry like SU(3) takes place for the above specific combination of
surface diffuseness and tetrahedral-type octupole deformation.  It
raises an interesting question on the relation between the
symmetry restoration and the tetrahedral deformation, and further
studies are necessary to clarify it.

\section{Nuclear prolate-shape dominance}
\label{sec:prodom}

Predominance of prolate shapes in nuclear ground-state deformation
(which is referred to as \textit{prolate dominance} for short) is a
long-standing problem of nuclear structure physics\cite{BM}.  Only a
few oblate ground states are found experimentally in medium to heavy
nuclei.  The microscopic mean-field theories also support this
feature\cite{Tajima1}.  In this section, we try to explain this
peculiar property of nuclei from the semiclassical point of view with
a realistic nuclear mean field model taking account of spin-orbit
coupling.

\subsection{Some earlier studies and remaining problems}

Various approaches have been attempted aiming at a simple
interpretation of the prolate dominance in nuclear ground-state
deformations.  It is generally recognized that the surface property of
the mean-field potential has relevance to the deformed shell structures
responsible for the prolate dominance.  The surface of the mean field
potential becomes sharper with increasing mass number.  The transition
of the deformed shell structure from light to heavy nuclei are studied
by Strutinsky \etal \cite{StrMOD} using the WS potential model with
spheroidal deformation.  The deformation is
parametrized by the axis ratio $\eta=R_z/R_\perp$ where $R_z$ and
$R_\perp(=R_x=R_y)$ are semiaxes of the nuclear surface.  They
calculated the shell energy $\delta E(N;\eta)$ as functions of
deformation $\eta$ and the particle number $N$, and investigated its
ridge-valley structures in the $(\eta,N)$ plane.  For the prolate
deformation ($\eta>1$), the shell energy valleys have positive slopes
in small $N$ region while they turn into negative slopes in large $N$
region.  Such transition has been successfully explained using the
POT.  For small $N$, the surface diffuseness $a$ is
comparable with the nuclear radius $R_A$ and the WS potential can be
approximated by the anisotropic HO potential (\ref{eq:h_ho}).
Imposing the volume conservation condition
$\omega_\perp^2\omega_z=\omega_0^3$, the oscillator frequencies are
given as functions of axis ratio $\eta=\omega_\perp/\omega_z$ by
\begin{subequations}
\begin{align}
\omega_\perp &= \omega_0\eta^{1/3},\\
\omega_z &= \omega_0\eta^{-2/3}.
\end{align}
\end{subequations}
In the normal deformation region, the two-parametric shortest
equatorial orbit family makes the dominant contribution to the
periodic-orbit sum.  Its action integral is expressed as
\begin{equation}
S_\beta(E;\eta)=\frac{2\pi E}{\omega_\perp(\eta)}
=\frac{2\pi E}{\omega_0\eta^{1/3}}.
\end{equation}
Then, the constant-action lines (\ref{eq:cac}) behave as
\begin{equation}
N(\propto E_F^3) \propto \eta
\end{equation}
and they have positive slopes in the $(\eta,N)$ plane.  On the other hand,
for large $N$, the surface diffuseness is much smaller than the
nuclear radius and the WS potential looks more like a
square-well potential, and it might be further approximated by the infinite-well
(cavity) potential.  In the spheroidal cavity, shortest equatorial
orbits form a one-parametric family, while the meridian-plane orbits form
a two-parametric family due to the specific symmetry of the system.
Therefore, the meridian orbit families (triangular, quadrangular, ...) make
dominant contributions to the PO sum.
Imposing the volume conservation condition $R_\perp^2R_z=R_0^3$, the
semi-axes $R_i$ of the equi-potential surface are given by
\begin{subequations}
\begin{gather}
R_\perp(=R_x=R_y)=R_0 \eta^{-1/3}, \\
R_z=R_0 \eta^{2/3},
\end{gather}
\end{subequations}
and the length $L_\beta$ of the meridian orbit, say, rhomboidal
orbit is estimated as
\begin{equation}
L_\beta=4\sqrt{R_\perp^2+R_z^2}=4R_0\eta^{-1/3}\sqrt{1+\eta^2}.
\end{equation}
Then the action integral is expressed as
\begin{equation}
S_\beta(p)=pL_\beta\propto pR_0\eta^{-1/3}\sqrt{1+\eta^2}
\end{equation}
and the constant action lines behave as
\begin{equation}
N(\propto p_F^3)\propto \frac{\eta}{(1+\eta^2)^{3/2}}, \label{eq:cac_cavity}
\end{equation}
which have negative slopes in the prolate region $\eta>1$.

The semiclassical analysis of spheroidal cavity has been thoroughly
worked out by Frisk \cite{Frisk} using the Berry-Tabor trace formula.
The quantum mechanical shell energies are successfully reproduced by
the semiclassical formula as the sum over PO
contributions.  He has remarked that the curves (\ref{eq:cac_cavity})
in the oblate region $\eta<1$ are rather flat, and correspondingly,
the shell energy valleys running along them are also flat.  Hence,
no significant shell-energy gains are expected with oblate
deformations as nucleon numbers deviated from the
spherical magic numbers.  This explains the mechanism
of the prolate dominance very nicely.

Hamamoto and Mottelson \cite{HM2009} have discussed the origin of
the prolate dominance from a different point of view.  They compared
the behaviors of the single-particle levels against deformation in the
cases of the HO and cavity (infinite well) potential models.  In
axially deformed HO potential, the degenerate levels at the spherical
shape fan out freely with increasing deformation on both prolate and
oblate sides.  This is because the shell oscillator number is a good
quantum number in the HO model and interactions between levels from
different major shells are absent.  On the other hand, there are
interactions between inter-shell levels in the cavity potential, and
they affect the way of level fannings.  Their behavior on the prolate
and oblate sides show obvious asymmetry: the fannings of levels in the
oblate side are considerably suppressed in comparison to the prolate
side.  Hamamoto and Mottelson have compared the effects of the
interactions between inter-shell levels on the prolate and oblate sides,
and clarified the reason why the level fannings show the above
asymmetric behaviors on the prolate and oblate sides.  The suppression
of level fannings in the oblate side might reduce the chance to
acquire a reasonable shell energy gain by oblate deformation.  They
considered the above asymmetry in level splittings as the origin of
the prolate dominance.

Still there remain some questions to be answered.
Firstly, from the semiclassical point of view, the deformations are
essentially determined by the shell energy in which only short
POs make contribution, and they are related to the gross shell
structures.  However, the way of level splittings might be
related to rather fine structures of levels.  One should consider this
aspect more carefully.  Secondly, as suggested by Tajima \etal
\cite{Tajima2,Takahara}, prolate dominance shows strong
correlation with the strength of spin-orbit coupling, as well as the
surface diffuseness of the potential.  The effect of spin-orbit
coupling has not been studied in the above works.  The argument of
\cite{HM2009} applies to the case of realistic spin-orbit coupling
where one finds the same kind of prolate-oblate asymmetry in the level
fannings, but it cannot explain the \textit{oscillation} in
the prolate dominance with varying spin-orbit strength which Tajima
\etal have found.  Let us consider these issues in the following part.

\subsection{Gross shell structures in the power-law potential models}

\begin{figure}
\begin{center}
\includegraphics[width=.85\linewidth]{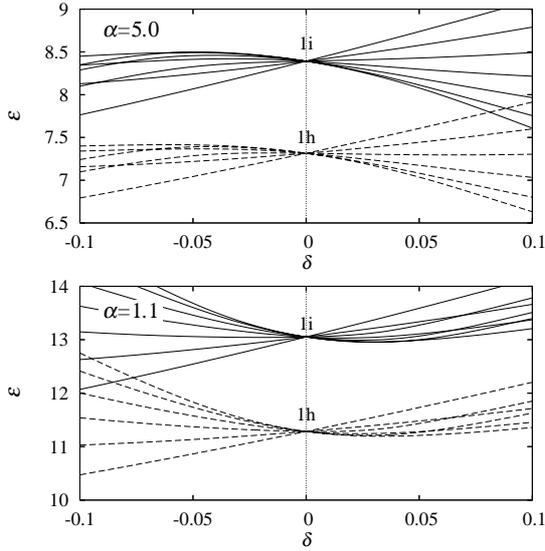}
\end{center}
\caption{\label{fig:fanning}
Splittings of the high-$j$ single-particle levels with spheroidal
deformation in power-law potential model for the power parameter
$\alpha=5.0$ and $1.1$.  Solid and dashed lines represent $1i$
and $1h$ levels, respectively.
Reproduced with permission from \cite{Arita2012}.
Copyright Americal Physical Society 2012.}
\end{figure}

First, let us generalize the above analyses to a model having a more
realistic radial dependence with finite diffuseness.  Spin-orbit
coupling is set aside for the moment.  In \cite{Arita2012}, we have
made analysis of the spherical and deformed shell structures for the
power-law potential $V\propto r^\alpha$ with varying the power
parameter $\alpha$.  For $\alpha=2$, corresponding to the HO
potential, level splitting occur in the same uniform way on both
prolate and oblate sides.  The shell energy valleys in the
$(\delta,N)$ plain have almost the same upward-right slopes on both
sides.  With increasing $\alpha$, the suppression of level fannings on
the oblate side manifests.  Figure~\ref{fig:fanning} shows the level
fannings of some high-$j$ levels.  In the upper panel for the power
parameter $\alpha=5.0$, corresponding to medium-mass nuclei, one
clearly sees a remarkable suppression of level fannings on the oblate
side.  In the lower panel, a calculation for $\alpha=1.1$ (although it
is an unrealistic value for any nucleus) is made, in which one finds a
suppression of level fannings on the prolate side, just as opposite to
the case of $\alpha>2$.

\begin{figure}
\begin{center}
\includegraphics[width=.8\linewidth]{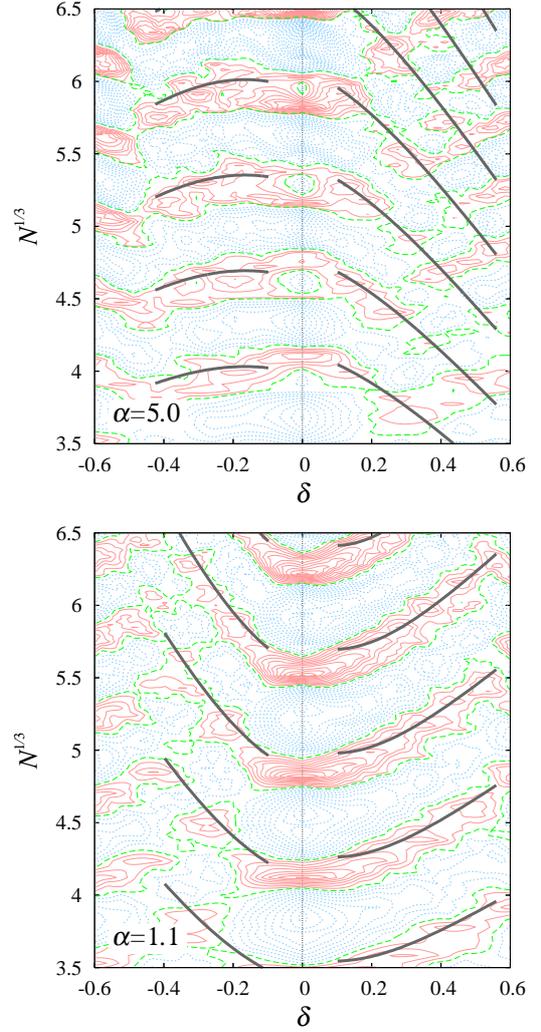}
\end{center}
\caption{\label{fig:scecont_nols}
Contour map of the shell energy $\delta E(N;\delta)$ in the
$(\delta,N^{1/3})$ plane for the spheroidal power-law potential model
without spin-orbit coupling.  Upper and lower panels are the results
for the power parameter $\alpha=5.0$ and $1.1$, respectively.  Solid
(red) and dashed (blue) contour lines are drawn for negative and
positive $\delta E$, respectively.  Thick lines represent the
constant-action ones (\ref{eq:cac}) of the bridge orbit C.}
\end{figure}

Figure~\ref{fig:scecont_nols} shows the contour maps of the shell
energy $\delta E(N;\delta)$ plotted as a function of deformation
$\delta$ and particle number $N$ for the above two values of the power
parameter $\alpha$.  Thick curves represent the constant action curves
(\ref{eq:cac}) of the POs which make dominant contribution to the
trace formula (\ref{eq:trace_esh}).  As we see in
figure~\ref{fig:bridge_nd}, the oval shape meridian-plane orbit
family C exists as the bridge orbit between equatorial diameter orbit
X and symmetry-axis diameter orbit Z.  This bridge orbit makes
significant contribution to the level density as expected from the
Fourier spectra in figure~\ref{fig:ftl_nd}.  In the upper panel of
figure~\ref{fig:scecont_nols} for $\alpha=5.0$, one finds that the
constant-action lines (\ref{eq:cac}) for the orbit C nicely explain
the shell energy valleys in the normal deformation region.  The slopes
of the curves are steep in the prolate side while they are rather flat
in the oblate side.  This behavior of the shell energy valleys
formed along the bridge orbit C can be considered as the
semiclassical origin of the prolate dominance.  Thus, one sees
that the argument of \cite{Frisk} for the spheroidal cavity model can be
generalized to more realistic potentials.

\begin{figure}
\begin{center}
\includegraphics[width=\linewidth]{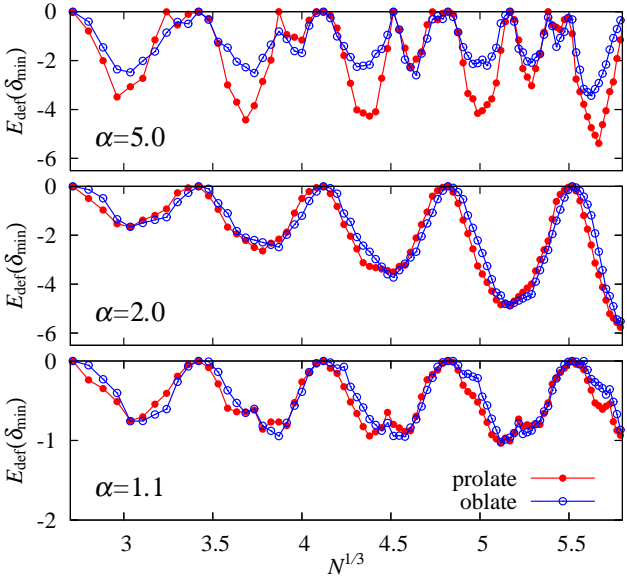}
\end{center}
\caption{\label{fig:emin_nols}
Comparison of the deformation energy minima (\ref{eq:edef}) on the
prolate ($\delta_{\rm min}>0$) and the oblate ($\delta_{\rm min}<0$)
sides, plotted as functions of the cubic root of the particle number $N$.
Bottom, middle and top panels show the results for the power parameter
$\alpha=1.1$, 2.0 and 5.0, respectively.}
\end{figure}

To examine if the prolate shapes are really favored in energy due to
the above behaviors of the deformed shell energies, we calculate the
ground-state deformations $\delta_{\rm min}$ by
minimizing the deformation energy
\begin{equation}
E_{\rm def}(N;\delta)=E(N;\delta)-E(N;\delta=0)
\label{eq:edef}
\end{equation}
with respect to the deformation parameter $\delta$ on each side of the
oblate ($\delta<0$) and prolate ($\delta>0$) shape, and compare the
energies at the prolate and oblate minima.  In evaluating the total
energy $E(N)$, sum of the single-particle energies $E_{\rm
sp}=\sum_{i=1}^N e_i$ is employed in \cite{HM2009}, but we shall make
a little improvement \cite{Arita2012}.  Writing the mean-field Hamiltonian as
$\hat{h}=\hat{t}+\hat{u}$ with $\hat{t}$ and $\hat{u}$ being the
kinetic energy and the mean-field potential, respectively, average
part of the total energy may be given by $\bar{E}\simeq
\<\hat{t}\>+\frac12\<\hat{u}\>$ if the mean field is a self-consistent
one from a certain two-body interaction.  Using
$\bar{E}_{\rm sp}=\<\hat{t}\>+\<\hat{u}\>$ and the Virial theorem
$\<\hat{t}\>=\frac12\<\br\cdot\bm{\nabla}\hat{u}\>
=\frac{\alpha}{2}\<\hat{u}\>$ for the power-law model, one has
\begin{subequations}
\begin{gather}
\<\hat{u}\>=\frac{2}{\alpha+2}\bar{E}_{\rm sp}, \\
\<\hat{t}\>=\frac{\alpha}{\alpha+2}\bar{E}_{\rm sp}, \\
\bar{E}=\frac{\alpha+1}{\alpha+2}\bar{E}_{\rm sp},
\end{gather}
\end{subequations}
and therefore
\begin{equation}
E(N)=\frac{\alpha+1}{\alpha+2}\bar{E}_{\rm sp}(N)+\delta E(N).
\end{equation}
Figure~\ref{fig:emin_nols} compares the deformation energy minima
$E_{\rm def}(\delta_{\rm min})$ on the prolate and oblate sides for
several values of the power parameter $\alpha$.  $\delta_{\rm min}$ is
the deformation parameter where the deformation energy (\ref{eq:edef})
takes minimum on each of the prolate and oblate side.  For
$\alpha=2.0$ (HO), systems with single-particle orbits filled up to
the lower half of the spherical shell prefer prolate shapes while
those up to the upper half
prefer oblate shapes, and the numbers of systems that have the lowest
energies at prolate and oblate shapes are
comparable.  For $\alpha=5.0$ where the potential surface is
considerably sharper than HO, prolate minima turn systematically lower
than oblate ones and they clearly show the prolate dominance.  For
$\alpha=1.1$, where one sees the suppression of level fannings on the
prolate side which may imply oblate-shape dominance, no essential
difference between prolate and oblate minima are found in the
deformation energies.  Looking at the shell energy contour plot for
$\alpha=1.1$ shown in the lower panel of
figure~\ref{fig:scecont_nols}, one sees that the constant-action lines
for the orbit C have considerably large slopes on the prolate side as
well as on the oblate side.  These constant-action lines nicely
explain the behavior of the shell-energy valleys, for which no
remarkable prolate-oblate asymmetry is expected.

To summarize above results, the prolate dominance is strongly correlated
with the behavior of shell energy valleys, and its origin
is clearly understood as the contribution of short classical POs.
Correlation between the level fannings and the shape dominance is missing
in the case of $\alpha=1.1$.  This might be because the level fannings
are related to rather finer shell structure associated with the
contribution of longer POs and their roles in the
shell energies (\ref{eq:trace_esh}) are less important.

\subsection{Effect of spin-orbit coupling}

Next, let us consider the effect of spin-orbit coupling.  By means of
the systematic Strutinsky calculations over the whole nuclear chart,
Tajima \etal examined the occurrence of the prolate dominance by
varying the surface diffuseness and spin-orbit strength of the
mean-field potential\cite{Tajima2,Takahara} in order to single out the
parameter which is playing the essential role.  They have calculated
the ground state deformations of all the observed combinations of
$(N,Z)$ in nuclear chart to extract the ratio of the numbers of
prolate and oblate ground states, and have examined its dependence on
the strength of $l^2$ potential (surface diffuseness) and $ls$
potential (spin-orbit strength) in the Nilsson (WS) model.  As the
results, they found a strong interference between the effects of
surface diffuseness and spin-orbit strength on the prolate/oblate
ratio.  Particularly, the prolate dominance disappears when the
spin-orbit parameter is reduced to the half of its realistic value.
Considering this result, the analysis based on the model without
taking account of spin-orbit coupling is giving us only partial
understandings for the prolate-shape dominance of real nuclei.  For a
deeper understanding of this feature, we make a semiclassical
analysis of the prolate-oblate asymmetry taking the spin-orbit
coupling into account.

\begin{figure}
\begin{center}
\includegraphics[width=.8\linewidth]{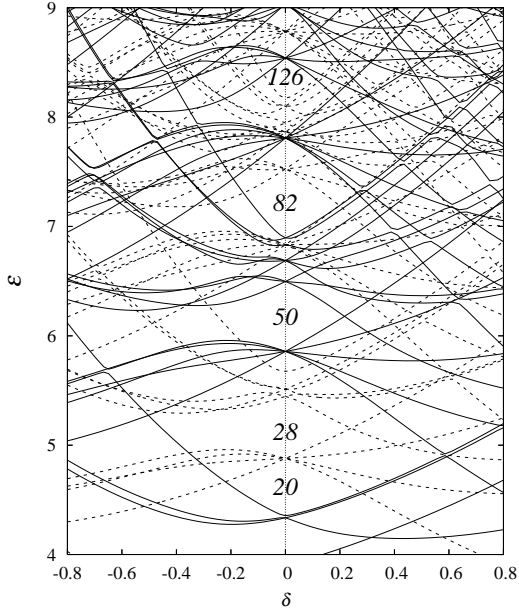}
\end{center}
\caption{\label{fig:nils_pw}
Single-particle level diagram for the power-law potential model with
the power parameter $\alpha=5.0$ and spin-orbit parameter $\kappa=0.06$.
Scaled-energy eigenvalues $\cE_j=(e_j/U_0)^{1/\alpha+1/2}$ are plotted
as functions of spheroidal deformation parameter $\delta$.  Solid and
broken lines represent positive and negative parity levels,
respectively.  The particle numbers of the closed-shell configurations
are indicated in italics.}
\end{figure}

Figure~\ref{fig:nils_pw} shows the level diagram for the power
parameter $\alpha=5.0$ and spin-orbit parameter $\kappa=0.06$, which
are considered as realistic for medium-mass nuclei.  The spherical
magic numbers (\ref{eq:magics}) are correctly reproduced with those
values of the parameters.  The behaviors of the level splittings with
increasing prolate and oblate sides look similar to the case without
spin-orbit coupling: one sees the same suppression of level fannings
on the oblate side as we see in the case without spin-orbit coupling
(see the upper panel of figure~\ref{fig:fanning}).  Hence it seems
that the argument in \cite{HM2009} also applies to the case of finite
spin-orbit coupling.

\begin{figure}
\begin{center}
\includegraphics[width=.8\linewidth]{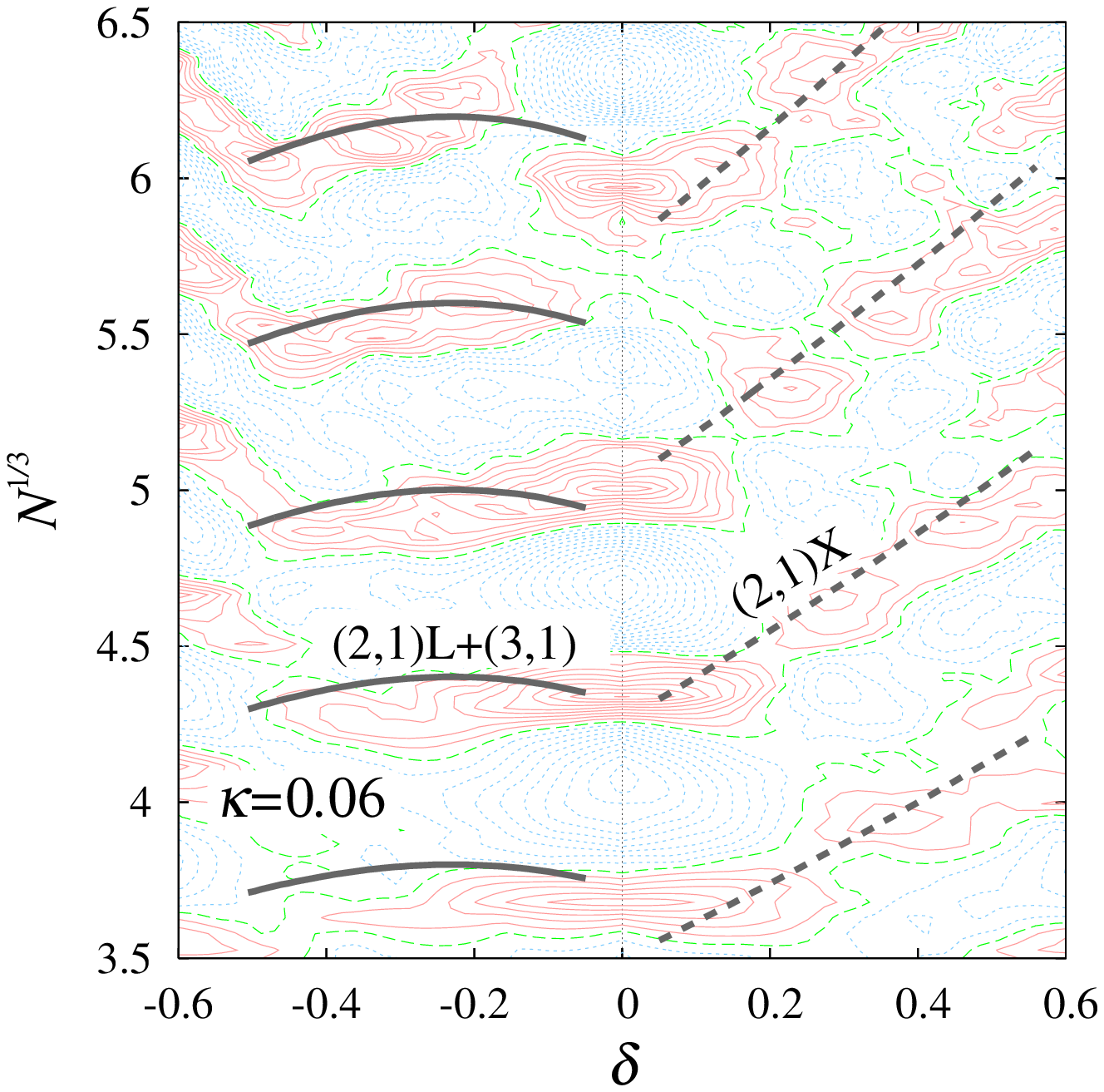}\\
\includegraphics[width=.8\linewidth]{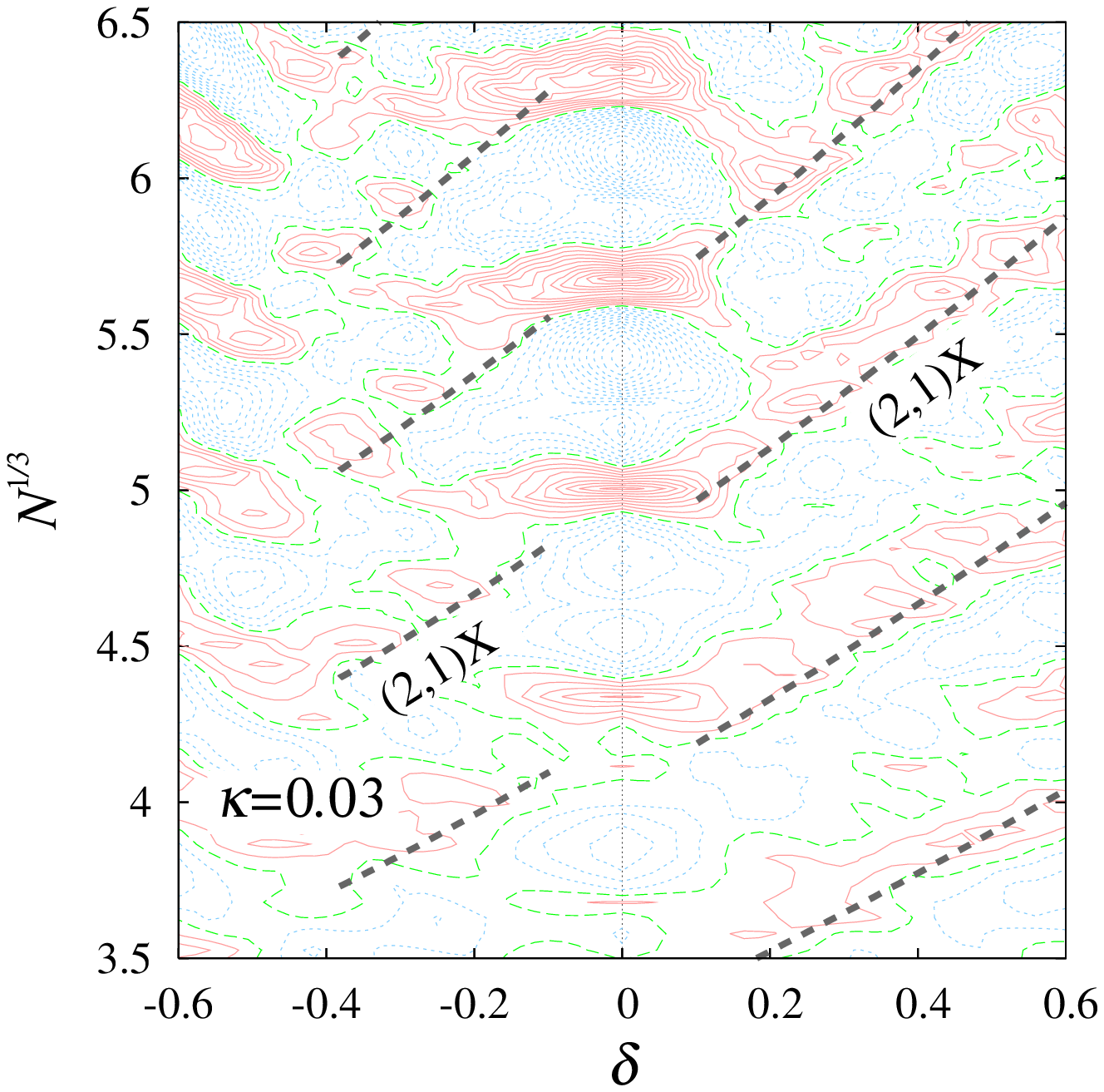}
\end{center}
\caption{\label{fig:scecont}
Contour map of the shell energy
$\delta E(N;\delta)$ in the $(\delta,N^{1/3})$
plane.  Solid (red) and dotted (blue) contour lines represent the
negative and positive $\delta E$, respectively.
Thick lines represent the constant-action ones (\ref{eq:cac}) for
some important short POs.}
\end{figure}

However, behavior of the deformed shell energies show quite strong
dependence on the spin-orbit parameter.  Figure~\ref{fig:scecont}
shows the contour maps of the shell energies $\delta E(N;\delta)$ as
functions of the deformation parameter $\delta$ and the particle
number $N$, for the power parameter $\alpha=5.0$.  We compare the
results for the case of realistic value of the spin-orbit parameter
$\kappa=0.06$ and for the reduced value $\kappa=0.03$ where Tajima
\etal found disappearance of the prolate dominance.
One may notice the obvious difference in the valley structures in the
deformed shell energies in those two maps, especially on the oblate
side.  For $\kappa=0.06$ the valley lines on the oblate side are
approximately flat, while for $\kappa=0.03$, one finds
valleys with considerably large slopes.

\begin{figure}
\begin{center}
\includegraphics[width=\linewidth]{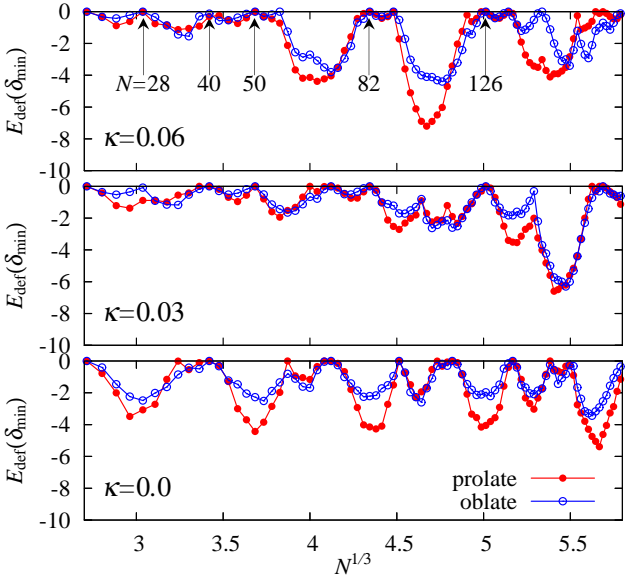}
\end{center}
\caption{\label{fig:emin}
The same as figure~\ref{fig:emin_nols} but for several values of the
spin-orbit parameter $\kappa$ with the fixed power parameter $\alpha=5.0$.
Bottom, middle and top panels show the results for $\kappa=0.0$,
0.03 and 0.06, respectively.}
\end{figure}

Figure~\ref{fig:emin} compares the prolate and oblate
deformation-energy minima $E_{\rm def}(N;\delta_{\rm min})$ for
different values of $\kappa$ with fixed value of the power parameter
$\alpha=5.0$.  The bottom panel for $\kappa=0$ is equivalent to the
top panel of figure~\ref{fig:emin_nols}.  With increasing $\kappa$,
the differences between prolate and oblate energy minima are reduced
at $\kappa=0.03$, the half of the realistic value, manifesting the
disappearance of prolate-shape dominance.  However, the differences
grow again for the realistic value $\kappa=0.06$ and the prolate
minima become considerably lower than the oblate ones, implying the
{\it revival} of the prolate-shape dominance.  All these results
nicely correspond to the behavior of shell energy valleys found in
figure~\ref{fig:scecont}.  Therefore, it is essential to describe the
above behavior of the shell energy valleys with varying spin-orbit
strength for understanding the origin of prolate dominance observed
in real nuclei.

\begin{figure}
\begin{center}
\includegraphics[width=\linewidth]{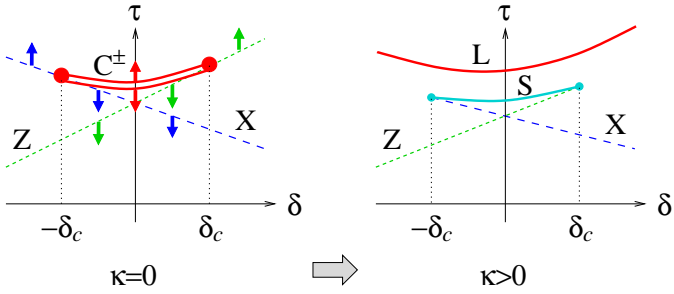}
\end{center}
\caption{\label{fig:bif21}
Illustration of the changes in the classical POs in the spheroidal
power-law potential model induced by the spin-orbit coupling.  The
scaled periods $\tau$ of some shortest POs are shown as functions of
the deformation parameter $\delta$.  Arrows in the left panel indicate the
directions of the changes in $\tau$ with increasing spin-orbit
strength $\kappa$.}
\end{figure}

To understand the above changes in deformed shell structures from
semiclassical view point, we investigated the effect of spin-orbit
coupling on the properties of the classical POs.
Figure~\ref{fig:bif21} illustrates what kinds of changes are induced
in the shortest POs when the spin-orbit coupling is
switched on.  For $\kappa=0$, one has two diameter orbits X and Z, and
the bridge orbits C connecting them at $\delta=\pm\delta_c$ as one
sees in figure~\ref{fig:bridge_nd}.  With increasing spin-orbit
strength $\kappa$, the periods of the orbits C$^\pm$, whose orbital
angular momenta parallel and anti-parallel to the spin, separate from
each other into L and S (denoting long and short, respectively).  The
changes in diameters X and Z show peculiar dependence on the
deformation.  The left part ($\delta<-\delta_c$) of X and the right
part ($\delta>\delta_c$) of Z changes into oval orbit with orbital
angular momenta parallel to the spin and are continuously connected
with L at $\pm\delta_c$, while the right part ($\delta>-\delta_c$) of
X and the left part ($\delta<\delta_c$) of Z changes into those having
the opposite directions of orbital angular momenta and cause tangent
bifurcations with S at $\delta=\pm\delta_c$.  One should also note
that, with increasing spin-orbit parameter $\kappa$, the bifurcation
deformation $\delta_c$ becomes smaller and the orbit (2,1)S shrinks to
a small deformation domain (see appendix for some detailed analyses on
those bifurcations).  On the other hand, the orbit (2,1)L survives for
any larger deformation.  With increasing $\kappa$, it undergoes
bifurcation and new triangular-type orbits (3,1)X and (3,1)Z (which
are symmetric with respect to the $x$ and $z$ axes, respectively) emerge
from it at around $\kappa=0.05\sim 0.06$, depending on the deformation
$\delta$.  Therefore, the orbits (2,1)L and (3,1)'s have almost the
same values of scaled periods at $\kappa=0.06$.  Those orbits should make
coherent contribution to the level density, and are expected to give
significant effects on the deformed shell structures for wide range of
deformation $\delta$ due to the bifurcation enhancement effect
discussed in section~\ref{sec:trace}.

\begin{figure}
\begin{center}
\includegraphics[width=\linewidth]{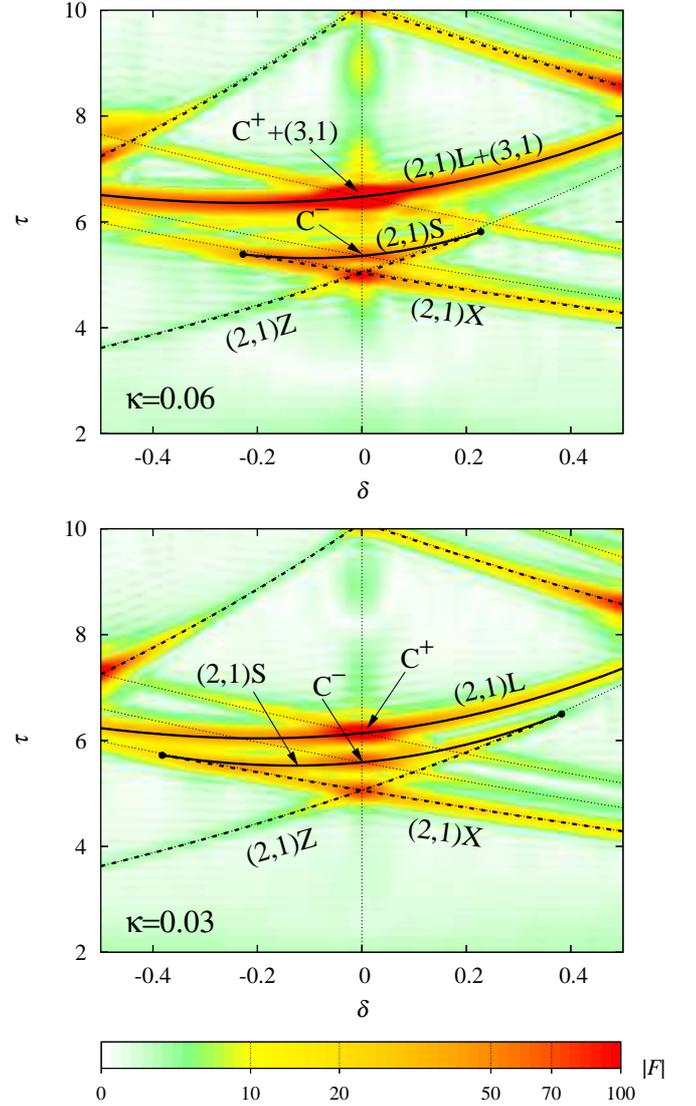}
\end{center}
\caption{\label{fig:ftd}
Color map of the quantum Fourier amplitude $|F^{\rm qm}(\tau;\delta)|$
in the $(\delta,\tau)$ plane.  The power parameter is $\alpha=5.0$ and
the spin-orbit parameter is $\kappa=0.06$ and $0.03$ for the upper and
lower panels, respectively.  Lines represent the scaled periods
$\tau_\beta(\delta)$ of the classical POs, and the dots indicate the
bifurcation points.}
\end{figure}

\begin{figure}
\begin{center}
\includegraphics[width=\linewidth]{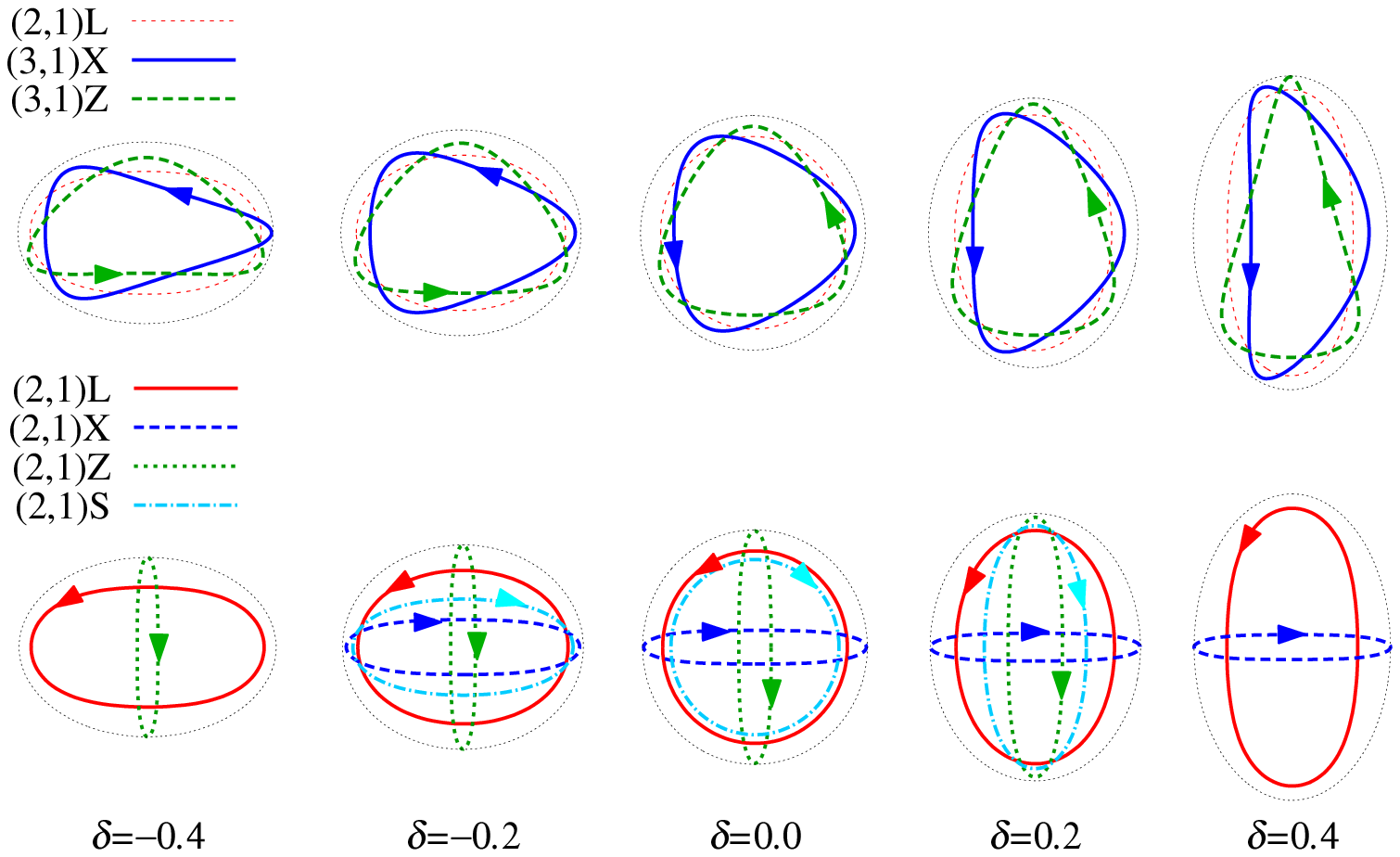}
\end{center}
\caption{\label{fig:po_d}
Some shortest meridian orbits for the power parameter $\alpha=5.0$ and
the spin-orbit parameter $\kappa=0.06$ with several values of the
deformation parameter $\delta$.  The orbit (2,1)S causes pair
annihilations with (2,1)X at $\delta=-0.23$, and with (2,1)Z at
$\delta=0.23$.}
\end{figure}

\begin{figure}
\begin{center}
\includegraphics[width=\linewidth]{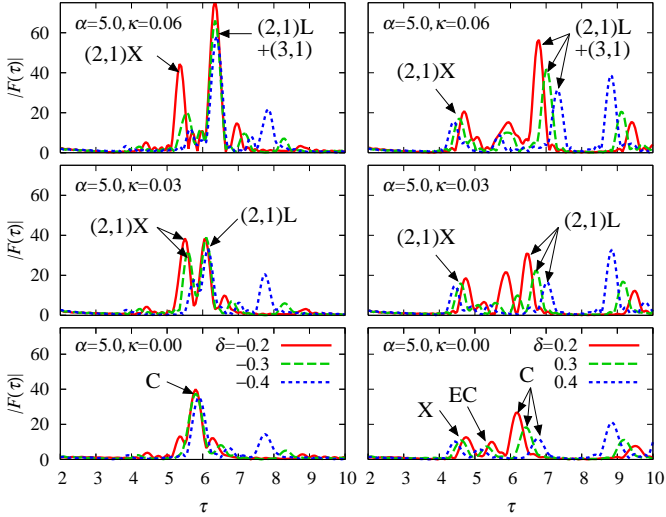}
\end{center}
\caption{\label{fig:fpeak_d}
Comparison of the quantum Fourier spectra $|F^{\rm qm}(\tau;\delta)|$
plotted for several values of the deformation parameter $\delta$ on
the oblate (left panels) and prolate (right panels) sides.  Bottom,
middle and top panels are the results for the spin-orbit parameter
$\kappa=0.00$, 0.03 and 0.06, respectively, with the power parameter
$\alpha=5.0$.}
\end{figure}

Thanks to the scaling relation for those frozen-spin orbits, we can
make use of the Fourier analysis in investigating how their
contributions change with varying $\kappa$.  Figure~\ref{fig:ftd}
shows the moduli of Fourier transforms $|F(\tau;\delta)|$ of the
quantum scaled-energy level densities, calculated for $\kappa=0.03$
and $0.06$.  Thick lines show the scaled periods of some shortest
POs which are displayed in figure~\ref{fig:po_d}.  One finds that the
Fourier amplitudes have peaks exactly along these meridian frozen-spin
POs.  The cross sections along the vertical line at several deformations
are shown in figure~\ref{fig:fpeak_d} in order to see the relative
strengths of the Fourier amplitudes.

On the prolate side, one finds considerable Fourier peak at the
shortest orbit (2,1)X, and it is expected to make major contribution
to the shell energy.  On the oblate side, the shortest orbit is
(2,1)Z, but its contribution is smaller than that of (2,1)X.  This is
because (2,1)Z occupies smaller phase-space volume since it is
isolated in the $\kappa\to 0$ limit.  Thus the second shortest (2,1)X
play major role also in the oblate side.  However, the orbit (2,1)X
reaches only up to $\delta\sim -0.2$ for $\kappa=0.06$, and it cannot
contribute much to the shell structures on the oblate side.  At this
realistic value of $\kappa$, the orbit (2,1)L causes a bifurcation
from which triangular-type orbits (3,1) emerge as discussed above.
With this bifurcation enhancement effect, their contribution to the
shell energy become dominant on the oblate side.  As one sees in the
upper panel of figure~\ref{fig:scecont}, the constant-action lines of
(2,1)L+(3,1) nicely explain the flat valleys in shell energy on the
oblate side.  For $\kappa=0.03$, the half of the realistic value, the shell
energy valleys are explained by the orbit (2,1)X on both prolate and
oblate sides.  Those valleys may play roles in establishing good
oblate and prolate minima, and explain the reason for the
disappearance of prolate dominance at this value of $\kappa$.  In this
way, the change in prolate-oblate asymmetry with varying spin-orbit
coupling can be clearly understood from the properties of the classical POs.

In summary, we consider the behavior of the shell energy valleys in the
$(\delta,N)$ plane which provides us the key to understand the
origin of prolate-shape dominance in nuclear
ground-state deformations.  These shell energy valleys have large
slopes on the prolate side while they are approximately flat in the oblate
side, and one has less possibility to acquire shell energy gains
with oblate deformations.  In practice, one sees nice correspondence
between the properties of the valley slopes and the deformed
shell-energy gains.  The features of the
level fannings should also have some effects but seems to bear less
important roles in the gross shell effects.
The way in which POs contribute to the shell energy is
quite sensitive to the spin-orbit parameter.  Although the shell
energy valleys for $\kappa=0$ look similar to the case of realistic
value $\kappa=0.06$, the semiclassical mechanisms for the enhancement
of the PO contributions are quite different in both cases.
Thus, above semiclassical interpretation gives us a deeper
understanding on the origin of prolate dominance for realistic
nuclear systems.

\section{Conclusions and perspectives}
\label{sec:summary}

Applying the semiclassical POT to the radial
power-law potential models, emergence of a rich variety of nuclear
shell structures are investigated from the view point of
quantum-classical correspondence.  In our semiclassical analyses, we
make full use of the scaling properties of the power-law potential
model and the Fourier transformation techniques, which are also
effective under the existence of spin-orbit coupling.  We have
emphasized the significant roles of the PO bifurcations for the
remarkable enhancement of shell effects with varying the parameters
like surface diffuseness (controlled by the power parameter $\alpha$),
deformations and spin-orbit coupling strength.  At the bifurcation
points, a family of quasi-periodic orbits appears around the
bifurcating PO, where an approximate dynamical symmetry is locally
restored.  In the bridge-orbit bifurcation, the above local family
occupies a large volume of the phase-space extending along the trail
of the bridge which connects two widely separated POs, and brings
about a larger dynamical symmetry compared to those for simple
bifurcations.  We have found such peculiar bridge-orbit bifurcations
play pivotal roles in exotic nuclear deformations such as
superdeformations and tetrahedral deformations.
It is also interesting to note that,
the SU(3) symmetry of the spherical HO Hamiltonian,
once broken for sharp potential surface with power parameter $\alpha>2$,
is partially compensated by the spin-orbit coupling as we see in
section~\ref{sec:pseudospin}, and also by the octupole deformation of
$Y_{32}$ type as we see in section~\ref{sec:octupole}.
Semiclassical analyses based on the realistic model for nuclear mean
field with spin-orbit coupling taken into account
provide us a deep understanding of the origin of the prolate-shape
dominance, which may show up under a delicate balance between the effect
from the surface diffuseness and that from the spin-orbit coupling.

Analyses of other types of nuclear deformations using the power-law
potential model with and without spin-orbit coupling are also
intended.  For instance, properties of shell structure under
reflection-asymmetric shapes by considering combinations of different
types of the octupole deformations, also with the quadrupole terms,
should be interesting, which might be responsible for
the systematics of ground-state deformations with reflection
asymmetry, and also for the fragment mass asymmetries in nuclear
fissions.

In this paper, we have used the semiclassical trace formula to extract
information on contributing classical POs from the quantum-mechanically
calculated single-particle spectra by means of the Fourier transformation
technique and the methods of
constant-action lines, but have not directly estimated the
semiclassical level densities and shell energies in the bifurcation region.
It is a challenging subject to develop
analytic and numerical methods to evaluate semiclassical trace
formula valid under existence of continuous symmetries, bifurcations,
and coupling with spin degree of freedom.  This becomes actually
important when we apply the semiclassical theories to more general
Hamiltonians without scaling, which would be required, for instance,
in descriptions of weekly bound nuclei.
Concerning the shell structures of such unstable nuclei, quenching
of the spherical shell gaps might be
one of the interesting subjects\cite{DHN94}.

\section*{Acknowledgments}
The author thanks K.~Matsuyanagi for discussions, comments and
careful reading of the manuscript.

\setcounter{equation}{0}
\renewcommand{\theequation}{A\arabic{equation}}
\section*{Appendix: Analyses of periodic orbit bifurcations with
monodromy matrices}

Classical PO changes their shapes continuously with varying potential
parameter.  The monodromy matrix (\ref{eq:m_monod}) varies accordingly
and one of the eigenvalues may coincides with unity which causes a
bifurcation the orbit.  Due to the symplectic property of the
Hamiltonian dynamics, the monodromy matrix $M$ is real and symplectic:
\begin{equation}
M^T J M=J, \quad J=\begin{pmatrix}0 & -I \\ I & 0\end{pmatrix}.
\end{equation}
Hence, the eigenvalues of $M$ always appear in a conjugate-reciprocal
pair either of
\begin{enumerate}\def\labelenumi{(\roman{enumi})}\itemsep=0pt
\item $\rme^{\pm\rmi v}$ with real $v$ (elliptic)
\item $\rme^{\pm u}$ with real $u$ (hyperbolic)
\item $-\rme^{\pm u}$ with real $u$ (hyperbolic with reflection)
\end{enumerate}
or in a quartet
\begin{enumerate}\def\labelenumi{(\roman{enumi})}
\setcounter{enumi}{3}\itemsep=0pt
\item $e^{\pm u \pm\rmi v}$ with real $u$ and $v$ (loxodromic).
\end{enumerate}
The bifurcation takes place at $v=0$ in the case (i) or at $u=0$ in
the case (ii).  The POs are stable if the monodromy matrix has only
elliptic eigenvalues, and otherwise unstable because the deviations of
the initial condition will grow exponentially as time evolves.

For the 2D systems, or the 3D systems with axial
symmetry, generic PO has a $(2\times 2)$ (symmetry reduced) monodromy
matrix and its eigenvalues appear in a pair either of (i)--(iii).  The
stability of a PO is uniquely determined by the value of the stability
factor $t=\Tr M-2=-\det(M-I)$ as
\begin{enumerate}\def\labelenumi{(\roman{enumi})}
\item $t=2\cos v-2=-4\sin^2(v/2)$,\quad $-4\leq t\leq 0$
\item $t=2\cosh u-2=4\sinh^2(u/2) > 0$
\item $t=-2\cosh u-2=-4\cosh^2(u/2) < -4$
\end{enumerate}
For 3D systems without continuous symmetry, generic PO
has $(4\times 4)$ monodromy matrix, and its four eigenvalues can be
generally expressed as
$(\lambda_1,\lambda_1^{-1},\lambda_2,\lambda_2^{-1})$.  Then the
stability of the PO is determined by the two stability factors $t_i$
$(i=1,2)$ defined by
\begin{equation}
t_i\equiv \lambda_i+\lambda_i^{-1}-2=-(\lambda_i-1)(\lambda_i^{-1}-1).
\end{equation}
With the relations
\begin{equation}
\Tr(M-I)=t_1+t_2, \quad \det(M-I)=t_1t_2,
\end{equation}
the above stability factors are simply obtained as the two roots of
the quadratic equation
\begin{equation}
t^2-t \Tr(M-I)+\det(M-I)=0.
\end{equation}
If the monodromy matrix has two pairs of eigenvalues either
in (i)--(iii) above, both $t_1$ and $t_2$ take the real values as in
2D cases,
while they take complex values $t_1, t_2=t_1^*$
for the loxodromic case (iv).

\begin{figure}
\begin{center}
\includegraphics[width=\linewidth]{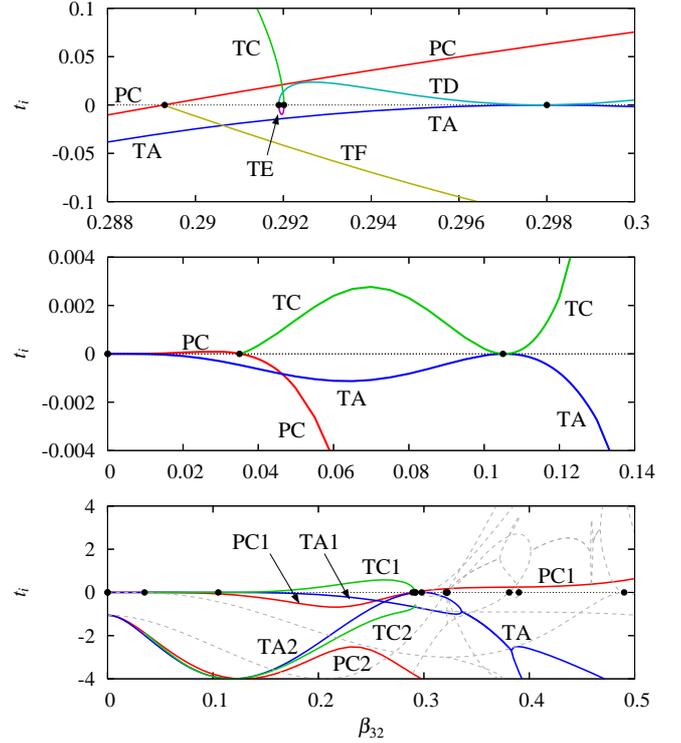}
\end{center}
\caption{\label{fig:trm_oct}
Stability factors $t_i$ (or $\Re t_i$ for loxodromic stability) of some
shortest POs as functions of the octupole deformation parameter
$\beta_{32}$.  Dots indicate the bifurcation points.}
\end{figure}

As the first example, we shall discuss some details on the
bifurcations found in the power-law potential with $Y_{32}$
deformation which we discussed in section~\ref{sec:octupole}.  The
planar triangular orbit PC, and 3D triangular orbits TA and TC (see
figure~\ref{fig:po_y32}) play the most important roles in deformed
shell structure.  Figure~\ref{fig:trm_oct} displays the values of
$t_i$ for some shortest POs as functions of the deformation parameter
$\beta_{32}$.  As one sees in the bottom panel, one of the $t_i$'s for
the above three dominant POs are very close to zero in
$0<\beta_{32}<0.3$, and those POs undergo bifurcations almost
simultaneously at $\beta_{32}\sim 0.3$, where deformed shell effect is
extremely enhanced.  In the middle panel, expanded plots of those POs
in the small $\beta_{32}$ region are shown.  The planar orbit PC
undergoes pitchfork bifurcation at $\beta_{32}=0.035$ and the 3D orbit
TC emerges.  TC and TA undergo touch-and-go bifurcation at
$\beta_{32}=0.105$.  In the top panel, expanded plots around the
bifurcation deformation $\beta_{32}\sim 0.29$ are shown.  The orbit PC
undergoes pitchfork bifurcation at $\beta_{32}=0.289$ and 3D orbit TF
emerges.  A pair of 3D orbits TD and TE emerge via tangent bifurcation
at $\beta_{32}=0.292$, and TE causes a pair annihilation with TC just
after its emergence.  The orbits TD and TA undergo touch-and-go
bifurcation at $\beta_{32}=0.298$.  As described here, the above
dominant POs are connected with each other by the complicated network
of quasi-periodic families of orbits via bifurcations.  This feature
seems peculiar to the tetrahedral-type deformations\cite{AriMuk}.

\begin{figure}
\begin{center}
\includegraphics[width=.9\linewidth]{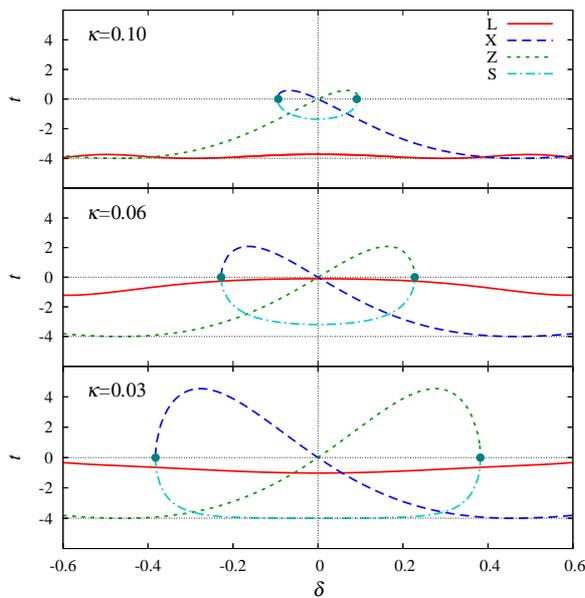}
\end{center}
\caption{\label{fig:trm_ls}
The stability factor $t$ of the frozen-spin meridian orbits
(2,1)L, (2,1)X, (2,1)Z and (2,1)S in spheroidal power-law potential
as functions of the deformation parameter $\delta$.  The results for the
power parameter $\alpha=5.0$ with three different values of the
spin-orbit parameter $\kappa=0.03$, 0.06 and 0.10 are shown.
Dots indicate the bifurcation points of (2,1)S.}
\end{figure}

For the second example, let us discuss the bifurcations of the orbits
(2,1) in spheroidal power-law potential with spin-orbit coupling which
we discussed in section~\ref{sec:prodom}.  We limit ourselves to the
frozen-spin orbits in the meridian plane, and only consider the
bifurcation within a given meridian plane, e.g. the $(x,z)$ plane.  Under
this limitation, bifurcations of POs can be considered in terms of the
reduced $(2\times 2)$ monodromy matrix $M$.  Figure~\ref{fig:trm_ls}
shows the values of the stability factor $t=\Tr(M-I)$ for four
oval-shape orbits (2,1) as functions of spheroidal deformation
parameter $\delta$.  For the spherical shape $\delta=0$, the orbits X
and Z compose the same degenerate family.  With increasing oblate
deformation $\delta<0$, orbit (2,1)X approaches (2,1)S and they
finally cause pair annihilation via tangent bifurcation at the certain
deformation $\delta=-\delta_c$.  With increasing prolate deformation
$\delta>0$, the orbit (2,1)Z approaches (2,1)S and they cause pair
annihilation via tangent bifurcation at $\delta=\delta_c$.  The value
of $\delta_c$ becomes smaller as increasing spin-orbit parameter
$\kappa$.  The value of $t$ for the orbit (2,1)L at $|\delta|\lesssim
0.2$ is very close to zero for $\kappa=0.06$ since the orbit is close
to the bifurcation point from which the orbits (3,1) emerge.  This
indicates that the (3,1) bifurcation makes significant effect on the
deformed shell structure for the realistic spin-orbit strength
$\kappa=0.06$ in rather wide range of the deformation parameter
$\delta$.

\end{document}